\documentclass[11pt]{article}
\usepackage[utf8]{inputenc}
\usepackage{diagbox}

\usepackage{algorithm}
\usepackage{algpseudocode}
\usepackage{amsmath}
\usepackage{amssymb}
\usepackage{mathtools}
\usepackage{bm}
\usepackage{amsthm}
\usepackage{mymaths}

\usepackage{enumitem}

\usepackage[font=small,labelfont=bf]{caption}

\usepackage{tikz-qtree}
\usepackage{tikz}

\usepackage[normalem]{ulem}

\title{
The Markov-Chain Polytope with Applications to  Binary  AIFV-$m$ Coding\footnote{Work of both authors partially supported by RGC CERG Grant 16212021.}}
   \author{
    Mordecai J. Golin  \hspace{0.25in}
    Albert John Lalim Patupat
}

\begin{document}

\maketitle

\begin{abstract}

This paper was motivated by the problem of constructing  cheapest binary AIFV-$m$ codes for a source alphabet of size $n.$
These
are a relatively new form of lossless codes that have better redundancy than Huffman codes,  achieved by using $m$ code trees rather than just one.
AIFV-$m$ coding  is   a special case of the problem of 
 finding a minimum-cost $m$-state Markov chain
$(S_0,\ldots,S_{m-1})$  in a large set of chains.

Each state has an associated  reward.
The cost of a chain is its {\em gain}, i.e., its  average  reward under its  stationary distribution.

Specifically, for each $k=0,\ldots,m-1,$ there is 
a known set 
$\mbbS_k$ of type-$k$ states. 
A chain is {\em permissible} if it contains exactly one state of each type; the problem is to find a minimum-cost permissible chain.

The problem of finding the minimum-cost  chain  models other coding problems as well.
The previously best known technique for  solving any of these problems  was an exponential-time iterative algorithm.

We describe  how to  map  type-$k$ states into  type-$k$ hyperplanes and then  define the  {\em Markov Chain Polytope} to be the lower envelope of all such hyperplanes.  
We then show that,  in many cases,  the Ellipsoid algorithm for linear programming run on this polytope can solve the minimum-cost Markov chain problem in polynomial time.

This paper is split into two separate parts.  The first describes the mapping that transforms the   generic minimum-cost Markov chain problem into a linear programming one.  The second specializes the first  to derive  the first polynomial time algorithm for constructing cheapest binary AIFV-$m$ codes for $m > 2.$ This  requires developing a deeper understanding of the combinatorics of AIFV codes.
\end{abstract}

 \section{Introduction}
 
\begin{figure}[t]
        \centering
        \includegraphics[width=4in]{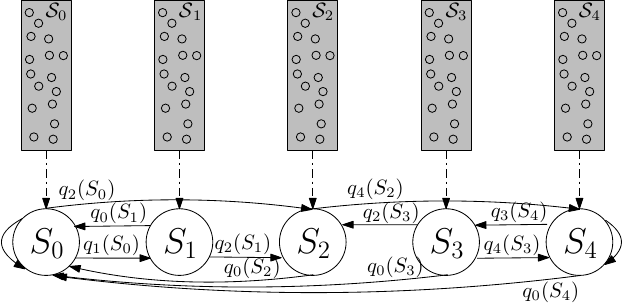}
        \caption{The bottom half of the figure illustrates   a five-state Markov chain.  Arrows represent non-zero transition probabilities.   $q_j(S_k)$ is the probability of transitioning from state  $S_k$ to  state $S_j.$ 
        Note that $q_0(S_k) >0$ for all $k\in [5].$  
        The circles in the shaded rectangle $\mbbS_k$ represent the set of all permissible type-$k$ states.
        Each  $S_k$ in the Markov Chain shown also belongs to the  associated set $\mbbS_k.$}
        \label{fig:MarkovEx}
        \end{figure}

 We  introduce the basic problem.
 Let $\bfS=(S_0,S_1,\ldots,S_{m-1})$ be an $m$-state Markov chain.   
 $[m]$ will denote the set 
$\{0,1,\ldots,m-1\}.$  See Figure \ref{fig:MarkovEx}.
 
 For $j,k \in [m],$ let  $q_j(S_k)$  be the  probability of transitioning from state $S_k$  to state $S_j$.  Assume further that 
 $\bfS$  has a unique stationary distribution 
$$\pii(\bfS) =(\pi_0(\bfS),\ldots,\pi_{m-1}(\bfS)).$$ 
Additionally, each state $S_k$ has an associated {\em reward} or {\em cost}  $\ell(S_k);$ the  {\em average steady-state cost}  or {\em gain} of $\bfS$  is  then defined   \cite{gallager2011discrete} as
$$\cost(\bfS) = \sum_{k=0}^{m-1} \ell(S_k) \pi_k(\bfS).
$$
A state $S_k$ is uniquely  
 defined by the $m+1$ values  $q_j(S_k), j \in [m]$ and $\ell(S_k).$

Next suppose that, for each $k \in [m],$ instead of 
there being only one $S_k,$ there exists a large  set  $\mbbS_k$  containing  all {\em permissible} ``type-$k$' states.  Fix 
$$\mbbS = \bigtimes_{k=0}^{m-1} \mbbS_k= \left\{(S_0,S_1,\ldots,S_{m-1}) \mid \forall k \in [m], S_k \in \mbbS_k\right\}$$ to be  their Cartesian product, a set of Markov chains, 
and further assume that $\forall \bfS \in \mbbS,$ $\bfS$ has a unique stationary distribution.
The problem is to find a  Markov chain  $\bfS \in\mbbS$  that minimizes $\cost(\bfS).$

This problem first arose  in the context of  binary AIFV-$m$ coding \cite{iterative_3,iterative_m,iterative_2} (also see   Section \ref{sec:AIFVlag}),
 in which  a {\em code}  is an $m$-tuple
$\bfT=(T_0,\ldots,T_{m-1})$  of binary  coding trees for a size $n$ source alphabet; for each $k \in [m],$ there are different restrictions on 
the structure of $T_k.$
The cost of  
 $\bfT$  is the  cost of  a corresponding  $m$-state Markov chain,  so the problem of finding the minimum-cost binary AIFV-$m$ code reduced to finding a minimum-cost Markov chain
\cite{iterative_4}.
  This same minimum-cost Markov chain approach  was later used to find better parsing trees 
\cite{iwata2021aivf},  lossless codes for finite channel coding  \cite{iwata2022joint} and AIFV codes for unequal bit-cost coding \cite{IHWY24}.

In   all  these problems, the input size $n$ was relatively small, e.g., a set of $n$ probabilities, but the associated $\mbbS_k$ had size exponential in $n.$
The  known algorithms  for solving the problem  were iterative ones 
that moved from Markov chain to Markov chain in $\mbbS$, in some non-increasing cost order. For the specific applications mentioned, they ran in exponential  (in $n$) time. Each iteration step also required solving  a local optimization procedure,  which was often polynomial (in $n$) time.

\cite{binary_search_2,GHIEEIT2023}  developed a different approach for solving the   binary  AIFV-$2$ coding problem, corresponding to a $2$-state Markov chain,  in  weakly polynomial time using a simple binary search. In those papers, they noted that they could alternatively  solve the problem in weakly polynomial time via the Ellipsoid algorithm for linear programming   \cite{ellipsoid}  on a two-dimensional   polygon.
They hypothesized that this latter technique could be extended to $m > 2$ but only with a better understanding of the geometry of the problem.

That is the goal of this work.
We first  define  a mapping of  {\em type-$k$ states}  to   {\em type-$k$ hyperplanes}  in $\RR^m$.  
We   show that  
the unique intersection of any $m$ hyperplanes, where each is of a different type,  always exists. We call such an intersection point ``distinctly-typed'' and prove that its ``height''  is equal to the  cost  of its  associated Markov chain.   The solution to the minimum-cost Markov-chain problem is thus  the lowest height of any  ``distinctly-typed' intersection point.  

We  then define the {\em Markov-Chain polytope}   $\MCP$ to be the lower envelope of  the hyperplanes associated with {\em all} possible  
states and prove  that some   lowest-height distinctly-typed intersection point is a highest point on $\MCP.$
We also show how, given {\em any} highest point on $\MCP$, to find a distinctly-typed intersection point on $\MCP$ of the same height.
This   transforms the very non-linear problem of finding the cheapest Markov chain  to the linear programming one  of finding a highest point of $\MCP.$

The construction and observations described above will be valid for ALL Markov chain problems.  
In the applications mentioned earlier, the  polytope    $\MCP$ is defined by an exponential number of constraints.  But, observed from the proper perspective, the local optimization procedures used at each step of the iterative algorithms in  \cite{iterative_3,iterative_m,iterative_2,iterative_4,iwata2021aivf}   can be repurposed  as polynomial time separation oracles for  $\MCP$.  This permits
 using the Ellipsoid algorithm approach of    \cite{ellipsoid} 
 to solve the binary AIFV-$m$ problem in weakly polynomial time instead of exponential time.


The remainder of the paper is divided into two distinct parts.  Part 1,  consisting  of Sections  \ref{sec:TopMC}-\ref{Sec:Proofs},  develops the general theory of transforming  the  minimum-cost Markov Chain problem into a linear programming one. Part 2, consisting of Sections  \ref{sec:AIFVlag}-\ref{sec:AIFVtech},
works out the details of how to apply Part 1's techniques  to construct minimum-cost binary AIFV-$m$ codes in weakly polynomial time.

More explictly, in Part 1, 
Section  \ref{sec:TopMC} discusses how to map the problem into a linear  programming one.  
Section \ref{sec:lemmas}  states our new results while Section \ref{Sec:AlgImp} discusses their algorithmic implications. In particular,  Lemma \ref{lem:newsol}  states sufficient conditions on $\mbbS$ that guarantee  a polynomial time algorithm for  finding the minimum cost Markov chain.
Section \ref{Sec:Proofs} then completes Part 1 by proving the main results stated in Section \ref{sec:lemmas}.
%
%
%

 Part 2,  in Sections  \ref{sec:AIFVlag}-\ref{sec:AIFVtech},
then discusses  how to apply Part 1's techniques  to construct best  binary AIFV-$m$ codes in weakly polynomial time.

Section \ref{sec:AIFVlag} provides necessary background, defining binary AIFV-$m$ codes and deriving their important properties.
Section \ref{sec:subAIFV} describes how to apply the techniques from Section \ref{Sec:AlgImp} to  binary AIFV-$m$ coding.
 Section \ref{sec:AIFVtech} proves a very technical lemma specific to  binary AIFV-$m$ coding required to show that its associated Markov Chain polytope has a polynomial time separation oracle, which is  the last piece needed to apply the Ellipsoid method.

Finally,  Section \ref{sec:newconc} concludes   with a quick discussion of other applications of the technique and possible directions for going further.

 We end  by noting that while this is the first polynomial time algorithm for constructing AIFV-$m$ codes for $m >2,$ it is only meant as a theoretical proof that  polynomial time algorithms exist.  Like most Ellipsoid based algorithms,  it would be difficult to implement efficiently in practice.




\section{Markov Chains}
\label{sec:TopMC}

\subsection{The Minimum-Cost Markov Chain problem}
\label{sec:MCMP}
\begin{definition}
\label{def:first}
Fix $m>1.$  
\begin{itemize}
\item[(a)] A {\em state} 
$S$ is defined by a  set of $m$  transition probabilities  $\{q_j(S)\}_{j \in [m]}$ along with a cost $\ell(S).$  $\forall {j \in [m]},\, q_j(S) \ge 0$ and $\sum_{j \in [m]} q_j(S)=1.$
\item[(b)] $\forall k \in [m],\,$ let $\mbbS_k$ be some  finite given set of states,
 satisfying that $\forall  S_k \in \mbbS_k,\,  q_0(S_k) >0.$  
The states in $\mbbS_k$ are known as {\em type-$k$} states.
\item[(c)] Markov Chain $ \bfS =(S_0,\ldots,S_{m-1})$ is  {\em permissible}  if $\forall k \in [m],\,  S_k \in \mbbS_k.$   
\item[(d)] Define  $\mbbS = \bigtimes_{k=0}^{m-1} \mbbS_k= 
\left\{
(S_0,\ldots,S_{m-1}) \mid \forall k \in [m],\, S_k \in \mbbS_k
\right\}$ to be the  set of permissible Markov chains.
\end{itemize}
\end{definition}
The actual composition and structure of each $\mbbS_k$ 
 is different from problem to problem and, within a fixed problem, different for different $k.$
The only universal constraint is 
(b),  that
 $\forall k \in [m],\,  \forall S_k \in \mbbS_k,$  $q_0(S_k) >0.$ This implies that 
 $\bfS=(S_0,\ldots,S_{m-1})$ is an ergodic {\em unichain},  with one aperiodic recurrent class  (containing $S_0$) and, possibly, some transient states. $\bfS$ therefore has a unique stationary distribution
  $\pi(\bfS),$
(where $\pi_k(\bfS)=0$ if and only if $S_k$ is a transient state.)   
\begin{definition}
\label{def:MCMC1}
Let $\bfS=(S_0,S_1,\ldots,S_{m-1})$  be a permissible   $m$-state Markov chain.  
The {\em average steady-state cost} of $\bfS$  is  defined to be
$$\cost(\bfS) = \sum_{k=0}^{m-1} \ell(S_k) \pi_k(\bfS).
$$
\end{definition}
This is  a {\em Markov Chain with rewards,} with  $\cost(\bfS)$ being its  {\em gain} 
\cite{gallager2011discrete}.

\begin{definition}
The {\em minimum-cost Markov chain problem}  is to find  $\bfS \in \mbbS$ satisfying
$$\cost(\bfS) = \min_{\bfS' \in \mbbS} \cost(\bfS').
$$
\end{definition}

{\em \par\noindent  Comments:\\
(i) In the applications motivating this problem,  each $\mbbS_k$ has size exponential in $n,$ so the search space $\mbbS$ has size exponential in $mn.$\\
(ii) 
The requirement in (b) that $\forall k \in [m],  q_0(S_k) >0,$ (which is satisfied by all the motivating applications) guarantees that $\bfS$ has a unique stationary distribution.
It is also used later in other places in the analysis.  Whether this condition is necessary is unknown; this is discussed  in Section \ref{sec:newconc}.
}

\subsection{Associated Hyperplanes and Polytopes}

The next set of definitions map type-$k$ states into type-$k$  hyperplanes in $\RR^m$ and then defines lower envelopes of those hyperplanes.  In what follows, $\xx$ denotes a vector $\xx=(x_1,\ldots,x_{m-1}) \in \RR^{m-1};$ $(\xx,y) \in \RR^m$ is a shorthand denoting that $\xx \in \RR^{m-1}$ and  $y \in \RR.$
$S_k$ denotes a state $S_k \in \mbbS_k.$ Recall that an  $S_k$ is uniquely defined by its $m$ transition probabilities and its cost.


\begin{definition} Let   $k \in [m]$.
\label{def:major1}
\begin{itemize}
\item  Define the type-$k$ hyperplanes $f_{k} : \RR^{m-1} \times \mbbS_{k} \to \RR$ as follows:
\begin{eqnarray*}
 f_{0}(\xx, S_0) &=& \ell(S_0) + \sum_{j=1}^{m-1} q_j(S_0) \cdot x_j,\\
\forall k >0,\ 
   f_{k}(\xx, S_k) &=& \ell(S_k) + \sum_{j=1}^{m-1} q_j(S_k) \cdot x_j - x_k.
\end{eqnarray*}
\item  For all $\xx \in \RR^{m-1},$ define  $g_{k} : \RR^{m-1} \to \RR$  and
$S_{k} : \RR^{m-1} \to\mbbS_{k}$
\begin{eqnarray}
g_{k}(\xx) &=& \min_{S_k \in \mbbS_{k}} f_{k}(\xx, S_k), \label{eq:g1}\\
 S_{k} (\xx) &=& \arg \min_{S_k \in \mbbS_{k}} f_{k}(\xx, S_k). \label{eq:g2}
\end{eqnarray}
Further,  define  the  Markov chain
$$\bfS(\xx)=  
\left(S_{0}(\xx),S_{1}(\xx) \ldots,S_{m-1} (\xx)
\right).
$$
For later use,  we note that, from Equations (\ref{eq:g1}) and  (\ref{eq:g2}),
\begin{equation}
\label{eq:fgcorr}
  \forall\xx \in \RR^{m-1},\, \forall k \in [m],\ 
  g_k(\xx) = f_k \left(  \xx,\, S_k(\xx)\right).
  \end{equation}

\item Finally,  for all $\xx \in \RR^{m-1},$  define.
$$
h(\xx) = \min_{k \in [m]} g_{k}(\xx)
=   \min_{k \in [m]}  \min_{S_k \in \mbbS_{k}} f_{k}(\xx, S_k).
$$
\end{itemize}
\end{definition}

$f_k$  maps  a  type-$k$ state $S_k$ to a type-$k$ hyperplane
 $f_{k}(\xx, S_k)$   in $\RR^{m}.$  For fixed $k,$  $g_{k}(\xx)$ is the 
{\em lower envelope} of all of the type-$k$ hyperplanes  $f_{k}(\xx, S_k).$  

Each $S_k(\xx)$ maps\footnote{Technically,  $S_k(\xx)$ is not {\em uniquely} defined for $\xx$ for which $g_k(\xx)$  is defined by the intersection of two or more hyperplanes. For those $\xx,$ $S_k(\xx)$ can be arbitrarily set to be any state $S_k\in \mbbS_{k}$ that achieves the $\arg\min.$ }  
  point $\xx \in \RR^{m-1}$ to  the lowest  
  type-$k$ hyperplane evaluated at $\xx.$  
$\bfS(\xx)$ maps  point $\xx \in \RR^{m-1}$ to  the  Markov chain $\bfS(\xx)\in \mbbS.$

 $h(\xx)$  will be the lower envelope of the $g_{k}(\xx)$.  
 Since  both  $g_{k}(\xx)$  and $h(\xx)$   are lower envelopes of hyperplanes,  they are the upper surface of convex polytopes in $ \RR^{m}.$ 
 This motivates: 
\begin{definition}
The {\em Markov Chain Polytope} in $\RR^m$ corresponding to $\mbbS$ is 
$$ \MCP = \left\{(\xx,y) \in \RR^{m} \mid  y \le h(\xx)\right\}.$$
\end{definition}

 \subsection{The Iterative Algorithm}
\label{subsec_iter}
  \cite{iterative_3,iterative_4,iterative_m,iterative_2}   present an iterative algorithm 
  that was first formulated for finding minimum-cost  binary AIFV-$m$ codes 
and then generalized  into a procedure for  finding  minimum-cost Markov Chains.

The algorithm starts with an arbitrary $\xx^{(0)}\in \RR^{m-1}$. It then iterates, at each step, constructing an $\xx^{(i)}\in \RR^{m-1}$ such that 
$\cost\left(\bfS\left( \xx^{(i)}\right) \right) \le \cost\left(\bfS\left( \xx^{(i-1)}\right) \right).$  The algorithm {\em terminates} at step $i$ if
$\xx^{(i)}=\xx^{(i-1)}.$

The fact that this algorithm terminates 
(Lemma \ref{lem:halting})  will be needed later (in Corollary \ref{cor:MCPtop2}) to prove that $\MCP$ contains some point corresponding to a minimum-cost Markov chain.

 \cite{iterative_2}  proves that the algorithm always terminates when 
 $m=2$ and, at termination, $\bfS\left( \xx^{(i)}\right)$ is a minimum-cost Markov Chain.
 \cite{iterative_3,iterative_4,iterative_m}   prove  that\footnote{They also claim  that the algorithm always terminates. The  proofs of correctness there are only sketches and  missing details but they all 
 seem to implictly assume that $q_j(S_k) >0$ for all $j,k \in [m]$  and not just $j=0.$
 The proof in \cite{DGZ1} only requires $q_0(S_k) >0.$ 
},
  for $m >2,$ if the algorithm terminates,  then  $\bfS\left( \xx^{(i)}\right)$  is a minimum-cost Markov Chain.
 A complete proof of termination  
in all cases  is provided in \cite{DGZ1}. This states
\begin{lemma}
\label{lem:halting}
[Theorem 1, \cite{DGZ1}]
For every 
 starting value $\xx^{(0)},$ there always exists $i$ such that $\xx^{(i)} =\xx^{(i-1)}.$ Furthermore, for that $i,$
\begin{equation}
\label{eq:eqg}
g_0\left(\xx^{(i)}\right) = g_1\left(\xx^{(i)}\right) = \cdots = g_{m-1}\left(\xx^{(i)}\right).
\end{equation}
\end{lemma}

Notes/Comments:
\begin{itemize}
\item The algorithm in  \cite{iterative_3,iterative_4,iterative_m,iterative_2}  looks different than the one described in \cite{DGZ1} but they are actually  identical, just expressed in  different coordinate systems.   The relationship between the two coordinate systems is shown in  \cite{golin2022speeding}.  The coordinate system in \cite{DGZ1} is the one used here.

\item Each step of the iterative algorithm requires  calculating $S_k(\xx^{(t)})$  for 
all $k \in [m].$
In applications, 
 finding  $S_k(\xx)$  is very problem specific and is usually 
  a combinatorial optimization problem. 
For example,  the first  papers on AIFV-$m$ coding \cite{iterative_2} and the most current papers on finite-state channel coding  \cite {iwata2022joint} 
calculate them using integer linear programming, as does a recent paper on constructing AIFV codes for unequal bit costs \cite{IHWY24}.
 The more recent papers on both AIFV-$m$ coding   \cite{dp_2,iterative_m,golin2022speeding}  and AIVF coding
 \cite{iwata2021aivf}
  use dynamic programming.
 \item The fact that if Equation (\ref{eq:eqg}) holds    
then $\bfS(\xx)$
 is a minimum-cost Markov Chain was proven directly in  \cite{iterative_3,iterative_4,iterative_m,iterative_2}.   An alternative proof of this fact is given in  Lemma \ref{lem:newopt}   in this paper.
 \item The proof of termination given in \cite{DGZ1} Theorem 1,  strongly depends upon condition (b) from Definition \ref{def:first}, i.e.,  $\forall k \in [m],\, q_0(S_k)>0.$
\end{itemize}

\section{The Main Results}
\label{sec:lemmas}
This section states our two main lemmas,  Lemma \ref{lem:geometric}   and  Lemma \ref{lem:Iterative_PruningX},  and their consequences.  Their proofs are deferred to Section \ref{Sec:Proofs}.
\begin{figure}[t]
\begin{minipage}{2.5in}
$
\begin{array}{|l||l|l|l|l||l|}
\hline
i & q_0(S_i) & q_1(S_i) & q_2(S_i) & \ell(S_i) \\ \hline \hline
0 & 0.5      & 0.25     & 0.25     & 9      \\ \hline
1 & 0.75     & 0        & 0.25     & 11     \\ \hline
2 & 0.75     & 0.25     & 0        & 14      \\ \hline
\end{array}
$
\end{minipage}
\begin{minipage}{3in}
        \includegraphics[width=3.3in]{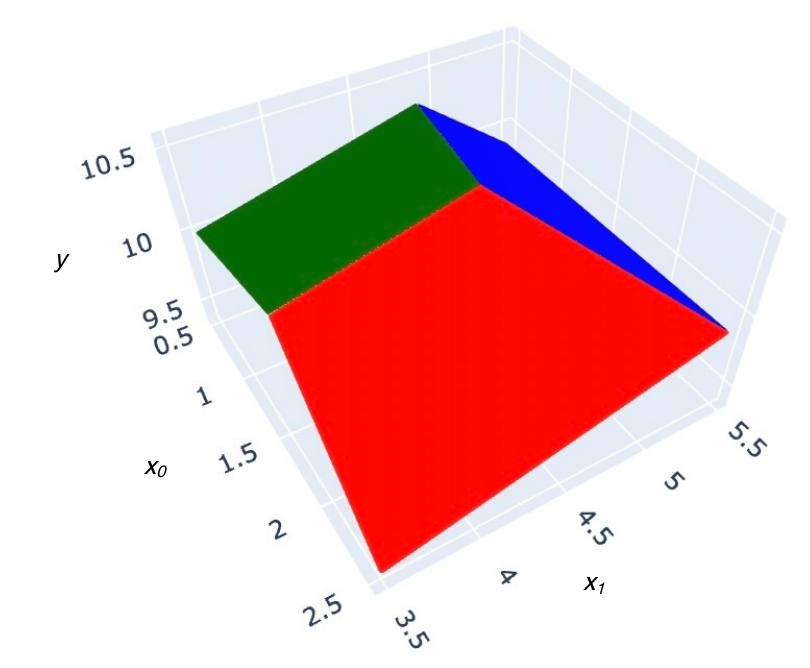}
\end{minipage}\caption{An illustration of Lemma \ref{lem:geometric}  (a) for a 3-state Markov chain $\bfS=(S_0,S_1,S_2).$ The table lists the associated $q_j(S_k)$ and $\ell(S_k)$ values. 
The green plane is 
$f_0(\xx,S_0)= 9 + x_1/4 + x_2/4,$
the  red plane  
$f_1(\xx,S_1)= 11 - x_1 + x_2/4,$
and the blue plane  
$f_2(\xx,S_2)= 14 + x_1/4 - x_2.$
By calculation,  $\pii(\bfS)=(0.6,0.15,0.25)$ so
$\mbox{\rm cost}(\bfS)= 0.6 \cdot 9 + 0.15\cdot 11 +0.25\cdot 14=10.55.$ The planes intersect at  unique point $(x_0, x_1, y)=(1.6,4.6,10.55)$.}
\label{fig:gen1}
\end{figure}

\begin{lemma}[Geometric Properties]
    \label{lem:geometric}   Let  $\bfS= (S_0,\ldots,S_{m-1}) \in\mbbS$ be any  permissible Markov Chain.
    \begin{itemize}
        \item[(a)] The $m$ $(m-1)$-dimensional hyperplanes   $y=f_{k}(\xx, S_k),$ $k \in [m],$
    intersect at a unique point $(\xx^{int}, y^{int})\in \RR^{m}$.\\
    We call such a $(\xx^{int}, y^{int})$ point a {\em distinctly-typed intersection point}.
        \item[(b)] $\forall \xx \in \RR^{m-1},\quad   y^{int} = \cost(\bfS)= \sum_{k=0}^{m-1} f_{k}(\xx, S_k) \cdot \pi_k(\bfS).$
        \item[(c)] The intersection point $(\xx^{int}, y^{int})$ is on or above the   lower envelope of 
        the $m$ hyperplanes $f_{k}(\xx, S_k),$ $k \in [m],$ i.e., 
        $$
        \forall \xx \in \RR^{m-1}, \quad y^{int} \geq \min_{k \in [m]} f_{k}(\xx, S_k).
        $$
        \item[(d)]      The intersection point $(\xx^{int}, y^{int})$ also satisfies  
        $$ \forall \xx \in \RR^{m-1}, \quad y^{int} \geq  h(\xx).$$
    \end{itemize}
\end{lemma}
Point (a) says that every permissible Markov chain defines a  distinctly-typed intersection point. 
The first part of point (b) states  that the  {\em height} ($m$-th coordinate) of that point is  the cost of the corresponding Markov chain,
immediately implying that 
finding  a minimum-cost Markov Chain is equivalent to finding a minimum height
distinctly-typed intersection point.  Parts (b) and (d) also imply  the  following corollary.
\begin{corollary}
\label{cor:minmax1}  $ \min_{\bfS \in \mbbS } \cost(\bfS)  \ge 
    \max_{\xx \in \RR^{m-1}} h(\xx).$
\end{corollary}

This in turn,  permits,  proving 
\begin{lemma}
\label{lem:newopt} 
If for some $\xx^* \in \RR^{m-1}$ and $y^* \in \RR,$
\begin{equation}
\label{eq:gh}
g_0(\xx^{*}) = g_1(\xx^{*}) = \cdots = g_{m-1}(\xx^{*})= y^*,
\end{equation}
 then
\begin{equation} 
\label{eq:newopt2}
y^*= h(\xx^*) = \cost\left(\bfS\left(\xx^*\right)\right)
\end{equation}
and
 \begin{equation}\label{eq:newopt3}
\min_{\bfS \in \mbbS } \cost(\bfS)   =   \cost(\bfS(\xx^*)) = h(\xx^*) = \max_{\xx \in \RR^{m-1}} h(\xx)=
\max\left\{
y \mid (\xx,y) \in \MCP
\right\}.
\end{equation}
\end{lemma}
\begin{proof}
By the definition of the $g_i$ and $h,$ if $\xx^*$ satisfies  (\ref{eq:gh}),  $h\left(\xx^*\right)= y^*.$

Recall that for all $k \in [m],$  $g_k(\xx) = f_k(\xx,S_k(\xx)).$
Thus,  (\ref{eq:gh}), implies that the hyperplanes $f_k(\xx,S_k(\xx^*))$ intersect at the point 
$\left(\xx^*,y^*\right)=\left(\xx^*,h\left( \xx^*\right)\right).$
Applying Lemma \ref{lem:geometric}  (b) proves (\ref{eq:newopt2}).

Combining  (\ref{eq:newopt2}) with  Corollary \ref{cor:minmax1}  immediately implies 
\begin{equation}
\label{eq:newopt4}
h(\xx^*)   =   \cost(\bfS(\xx^*))    \ge  \min_{\bfS \in \mbbS } \cost(\bfS)   \ge  \max_{\xx \in \RR^{m-1}} h(\xx)     \ge h(\xx^*).
\end{equation}
Because the leftmost and rightmost values in  (\ref{eq:newopt4})  are identical, the two inequalities in  (\ref{eq:newopt4})  must be equalities, proving (\ref{eq:newopt3}).
\end{proof}
%
%
%
Condition (\ref{eq:gh}) in Lemma \ref{lem:newopt} 
means  that the  $m$ different lower envelopes $g_k(\xx)$, $k \in [m],$  must simultaneously intersect at a point $\xx^*$.
 It is not a-priori obvious that such an  $\xx^*$ should always exist  
 but 
Lemma \ref{lem:halting}  tells us that the iterative algorithm always terminates at such an $\xx^*$,  immediately proving\footnote{For completness, we note that
Lemma \ref{lem:AIFVmax} later also {\em directly}  proves that condition  (\ref{eq:gh}) in Lemma \ref{lem:newopt} holds for the special case of AIFV-$m$ coding.  So, this paper provides a 
fully self-contained proof of Corollary \ref{cor:MCPtop2} for AIFV-$m$ coding without requiring  Lemma \ref{lem:halting}.}
\begin{corollary}\label{cor:MCPtop2}
There exists $\xx^* \in \RR^{m-1}$ satisfying  (\ref{eq:gh}) in Lemma \ref{lem:newopt}.  This $\xx^*$ satisfies
$\left(\xx^*,h\left(\xx^*\right)\right) \in \MCP$ and 
$h\left(\xx^*\right)=
\max\left\{
y \mid (\xx,y) \in \MCP
\right\}.$
\end{corollary}


\begin{figure}[t]
\centering
        \includegraphics[width=3in]{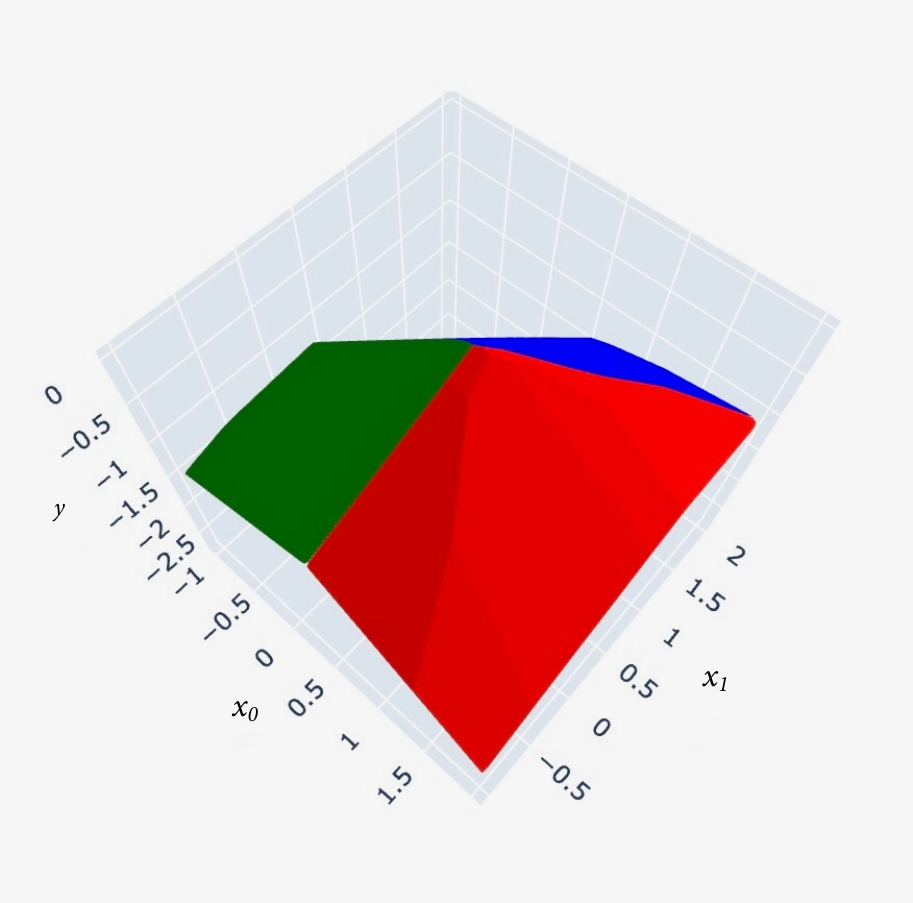}
\caption{An illustration of Lemma \ref{lem:newopt}  and Corollary \ref{cor:MCPtop2} for the case $m=3.$  3000 states each of type $0$, $1$ and $2$ were generated with associated $\ell(S_k)$ values. The green, red and blue  are, respectively the lower envelopes $g_0(\xx)$, $g_1(\xx)$, $g_2(\xx),$ i.e., the lower envelopes of the 3000 associated hyperplanes of each type.  $\MCP$ is the lower envelope of those three lower envelopes. The three $g_i(\xx)$  intersect at a unique point $(x_1^*,x_2^*,y^*)$ which  is a highest point in $\MCP.$  $\bfS^*=\bfS((x_1^*,x_2^*))$ is a minimal cost Markov chain among the  $3000^3$ permissible Markov chains in $\mbbS$ and $\cost(\bfS^*)= y^*$.}
\label{fig:gen2}
\end{figure}

This suggests  a new approach to solving our problem.
\begin{enumerate}
\item Use linear programming to  find  a highest point $(\hat \xx, \hat y) \in \mathbb H.$
\item Starting from  
$(\hat \xx, \hat y)$   find  {\em distinctly-typed} intersection point  $(\hat \xx^*, \hat y) \in \mathbb H.$
\item Return $\bfS(\xx^*).$
\end{enumerate}

$S(\xx^*)$ is a minimum-cost Markov chain, so this would solve the problem.
The complication is how to get from an   arbitrary highest {\em  point}  $(\hat \xx,\hat y) \in \MCP,$ to a distinctly-typed highest point  of $ \MCP$ (the solution to the problem).

%
%

This  complication can be sidestepped using  the  extension to $P$-restricted search spaces  described below.
The intuition is that  if  $h(\hat \xx)$ is a highest point on $\MCP$ but is not a distinctly-typed intersection  then some of its component states must be transient.  ``Pruning'' away these transient states will permit constructing  a mininimum-cost Markov Chain on the subproblem restricted to the non-transient states.  We note that the the intersection point  corresponding to this Markov-chain will NOT necessarily be on  $\MCP$.


\begin{definition}[$P$-restricted search spaces] 
Let 
\label{def:P_RestrictedX}$\calP$  denote the set of all subsets of  $[m]$ that contain  ``0'',  i.e., 
$\calP=\{ P\subseteq [m] \mid 0 \in P\}.$
For all $k\in [m]$ and all $s_k \in \mbbS_k,$ 
define $P(S_k) = \{j \in [m] \,:\, q_j(S_k) >0\},$ the set of all states to which $S_k$ can transition.   Since $q_0(S_k) >0,$  $0    \in P(S_k)$, so  $P(S_k) \in \calP.$

Now  fix  $k \in[m]$ and $P \in \PP.$
 Define 
$$\mbbS_{k|P} = \Bigl\{ S_k \in\mbbS_{k}  \mid   P(S_k) \subseteq P\Bigr\}.$$
to be the subset of states in $\mbbS_k$ that only transition to states in $P.$

 Further define 
$$\mbbS_{|P} = \bigtimes_{k=0}^{m-1} \mbbS_{k|P}.$$
Note that    $\mbbS_{k|[m]} = \mbbS_{k}$  and  $\mbbS_{|[m]}= \mbbS$. Also note that if $(S_0,\ldots,S_{m-1}) =\bfS \in\mbbS_{|P}$ then, if $ k \not\in P,$ $S_k$ is a transient state in $\bfS.$
\end{definition}

 \begin{definition} Let   $k \in [m]$ and $P \in \PP.$ 
 
\label{def:major2}
  For all $\xx \in \RR^{m-1},$ define  $g_{k|P} : \RR^{m-1} \to \RR$  and
$S_{k|P} : \RR^{m-1} \to\mbbS_{k|P}$ as
\begin{eqnarray*}
g_{k|P}(\xx) &=& \min_{S_k \in \mbbS_{k|P}} f_{k}(\xx, S_k),\\
 S_{k|P} (\xx) &=& \arg \min_{S_k \in \mbbS_{k|P}} f_{k}(\xx, S_k),
\end{eqnarray*}
and
$$\bfS_{|P} (\xx)=  
\left(S_{0|P}(\xx),S_{1|P}(\xx), \ldots,S_{m-1|P} (\xx)
\right).
$$
Note that  $g_{k}(\xx)$,  $S_{k} (\xx)$ and $\bfS (\xx)$ are, respectively,  equivalent to 
$g_{k|[m]}(\xx)$,  $S_{k|[m]} (\xx)$ and $\bfS_{|[m]} (\xx),$ i.e, all types permitted.
\end{definition}

 These definitions permit moving from {\em any} ``highest'' point on   $\MCP$  to a distinctly-typed intersection point at the same height (proof  deferred to Section \ref{Sec:Proofs}).
\begin{lemma} 
\label{lem:Iterative_PruningX} 
    Suppose  $(\hat \xx, \hat y) \in \mathbb H$ satisfies
     $\hat y =   \max \left\{ y \,:\, (\xx,y)\in \MCP  \right\}.$
     
Let $\bfS^*=(S^*_0,, \dots, S^*_{m-1})$ denote any minimum-cost Markov chain
 and 
 $P^* = \{ k : \pi_k(\bfS^*)> 0 \},$  the set of its recurrent indices $k$.  
Let $P \in \calP.$
   Then
    \begin{itemize}
        \item[(a)] 
        If $\forall k \in P,$ $g_{k|P}(\hat \xx) = \hat y$, then 
$\cost\left(\bfS_{|P}(\hat \xx)\right)= \hat y.$\\ Equivalently,  this implies $\bfS_{|P}(\hat \xx)$
 is a  minimum-cost Markov chain.
        \item[(b)] 
       If $P^* \subseteq P$, then 
       $\{0\} \subseteq P^* \subseteq \{ k\in P : g_{k|P}(\hat \xx) = \hat y \}$. 
    \end{itemize}
\end{lemma}

\begin{corollary}[Iterative Pruning Yields Optimal Solution] 
\label{cor:IterativeAlg}
 \ \\
    Suppose $(\hat \xx, \hat y) \in \MCP$ satisfies
     $\hat y =   \max \left\{ y \,:\, (\xx,y)\in  {\mathbb H} \right\}.$


%
%
   Then  the   procedure $\mbox{\rm Prune}(\xx)$ in Algorithm \ref {alg:prune} terminates  in at most $m-1$ steps. At  termination,  $\bfS_{|R}(\hat \xx)$ is a minimum-cost Markov chain.
    \begin{algorithm} 
    \begin{algorithmic}[1]
\caption{$\mbox{Prune}(\hat \xx):$   $(\hat \xx, \hat y) \in {\mathbb H}$ where    $\hat y =   \max \left\{ y \,:\, (\xx,y)\in  {\mathbb H} \right\}.$}
\State {Set $R  = [m]$}
\State {Set $R' = \{k \in  [m] \mid  g_{k|[m]}(\hat \xx) = \hat y \}.$}
\While {$R' \not=R$}
\State {Set $R  = R'$}
\State {Set $R' = \{k \in  R \mid  g_{k|R}(\hat \xx) = \hat y \}.$}
\EndWhile 
\label{alg:prune}
\end{algorithmic}
\end{algorithm}
\end{corollary}
\begin{proof} 
At the start of the algorithm,  $P^*  \subseteq  [m]=R.$ 
As long as $R' \not = R$,  Lemma \ref{lem:Iterative_PruningX} (b) implies that
$P^*  \subseteq  R' \subseteq R.$   If $R \not=R'$,  the size of $R$ decreases, so  the process can not run more than $m-1$ steps.  At termination, $R'=R,$ so from  Lemma \ref{lem:Iterative_PruningX} (a),  $\bfS_{|R}(\hat \xx)$ is a minimum-cost Markov chain.
 \end{proof}

%
%
%

\section{Algorithmic Implications}
\label{Sec:AlgImp}

The previous section described how to  recast the minimimum-cost Markov chain problem into a linear programming problem with one constraint for every possible state,  along a special procedure at the end that transforms any solution to the linear program into one for the Markov chain problem.
This  section describes how to extend this formulation into one that can be solved using the Ellipsoid algorithm for linear programming.

Doing so  requires introducing a few  further definitions.
\begin{definition}
Let ${\bf R} = \bigtimes_{1 \le k < m} [\ell_k,r_k]$ be a hyperrectangle in $\RR^{m-1}.$  Define
\begin{itemize}
\item $t_{\mbbS}({\bf R})  $ is the maximum time required to calculate $\bfS(\xx)$ for any $\xx\in {\bf R}.$
\item $t'_{\mbbS}({\bf R})  $ is the maximum time required to calculate $ \bfS_{|P}(\xx)$ 
for any $\xx\in {\bf R} $ and any $P \in \calP.$ 
\end{itemize}


 Note that for any ${\bf R},$   $t_{\mbbS}({\bf R})  \le t'_{\mbbS}({\bf R}).$
 Also,  $t_{\mbbS} $ and  $t'_{\mbbS} $  will denote  $t_{\mbbS}\left(\RR^{m-1}\right)$ and  
 $t'_{\mbbS}\left(\RR^{m-1}\right).$
 
 In addition, 
 \begin{itemize}
 \item Let $\mathcal I$ denote the number of iterative  steps made by the iterative algorithm.  (The only known bound for this is $|{\mathcal I}| \le |\mbbS|,$ the number of permissible Markov Chains.)
 \end{itemize}
\end{definition}
%
%

Since the iterative algorithm calculates a $\bfS(\xx)$ at every step, 
it requires $O( {\mathcal I}  \cdot t_{\mbbS})$ time. Improvements to its  running time have focused on improving $t_{\mbbS}$ in specific applications.

For binary AIFV-$m$ coding, of $n$ source-code words,  $\bfS(\xx)$  was first solved  using integer linear  programming \cite{iterative_2} so $t_{\mbbS}$ was exponential in $n.$
For the specific case of ${\bf R}=[0,1]^{m-1},$ this was improved  to polynomial time using different dynamic programs.   More specifically,  for $m=2,$   \cite{dp_2} showed that
$t_{\mbbS}({\bf R})=O(n^5)$, improved to $O(n^3)$ by \cite{dp_2_speedup}; for $m >2,$  \cite{iterative_m}
 showed that
$t_{\mbbS}({\bf R})=O(n^{2m+1})$, improved to $O(n^{m+2})$ by \cite{golin2022speeding}.  These sped up the running time of the iterative algorithm under the (unproven) assumption that the iterative algorithm always stayed within $[0,1]^{m-1}$.

For AIVF codes,    \cite{iwata2021aivf} proposed using a modification of a dynamic programming  algorithm  due to 
\cite{dube2018individually}, yielding that for any fixed $m$,  $t_{\mbbS}$ is polynomial time in $D,$ the number of words permitted in the parse dictionary.

In all of these  cases, though, the running time of the algorithm was  still exponential because ${\mathcal I}$
could  be exponential.

{\em Note: Although   $t_{\mbbS}({\bf R})$  has been  studied,  nothing was previously known about $t'_{\mbbS}({\bf R}).$ This is simply because there was no previous need to define and construct $ \bfS_{|P}(\xx)$. In all the known cases  in which algorithms for constructing  $\bfS(\xx)$ exist, it is  easy to slightly modify them to construct 
$\bfS_{|P}(\xx)$ in the same running time.  So $t'_{\mbbS}({\bf R}) = \Theta(t_{\mbbS}({\bf R})).$
}

We now see how the properties of the Markov chain polytope will, under some fairly loose conditions, permit  finding the minimum-cost Markov chain  in polynomial time. This will be done via 
 the Ellipsoid method of Gr{\"o}tschel,  Lov{\'a}sz and Schrijver \cite{ellipsoid,grotschel2012geometric,schrijver1998theory}, which, given a polynomial time {\em separation oracle}  for a polytope,  
 permits solving a linear programming problem on the polytope in polynomial time. The main observation is that $\bfS(\xx)$ provides a separation oracle for $\MCP,$  so the 
previous application-specific algorithms for  constructing   $\bfS(\xx)$ can be reused to derive polynomial time algorithms.


\subsection{Separation oracles and $\MCP$}

 Recall the definition of a {\em separation oracle}.
\begin{definition}[\cite{schrijver1998theory}]
Let  $K \subseteq  \mathbb{R}^m$ be a closed convex set.  A {\em separation oracle}\footnote{Some references label this  a {\em strong} separation oracle.  We follow   
 \cite{schrijver1998theory} in not adding the word {\em strong}.}
for $K$ is  a procedure that, for any
$\xx \in  \mathbb{R}^m,$ either reports that $\xx \in K$ or, if $\xx \not \in K$, returns  a hyperplane that separates $\xx$ from $K.$   That is, it returns $\bfa \in \mathbb{R}^m$ such that  $\forall \zz  \in K,$ ${\bfa}^T \xx > \bfa^T  \zz.$
\end{definition}

It is now clear that $\bfS(\xx)$ provides a separation oracle for $\MCP.$
\begin{lemma}
\label{lem:SSep} Let $m$  be fixed and 
$\MCP$ be the Markov Chain polytope.  Let $\zz = (\xx,y) \in \RR^m$.  Then 
 knowing $\bfS(\xx)=(S_0(\xx),\dots,S_{m-1}(\xx))$ provides a $O(m^2)$ time algorithm for either reporting that $\zz \in \MCP$ or returning  a hyperplane  that separates $\zz$ from $\MCP.$  
\end{lemma}
\begin{proof}
$\zz \in \MCP$ if and only if
$$y \le h(\xx) = \min_{k \in[ m]}  \left( \min_{S_k \in \mbbS_k} f_k(\xx,S_k)\right)
=\min_{k \in[ m]} g_k(\xx)
=\min_{k \in[ m]} f_k\left(\xx, S_k(\xx)\right).
$$
Thus knowing  $\bfS(\xx)$ immediately determines whether $\zz\in \MCP$ or not.  Furthermore if $\zz \not\in \MCP$, i.e., $y' > h(\xx),$  let $k'\in[m]$ be an index
satisfying
$f_{k'}\left(\xx, S_{k'}(\xx)\right)=h(\xx).$ 

The hyperplane $y=f_{k'}(\xx, S_{k'}(\xx))$ then separates  $(\xx, y')$ from $\MCP$ because $y=f_{k'}(\xx, S_{k'}(\xx))$ is a {\em supporting} hyperplane of $\MCP$  at point $(\xx,h(\xx)).$
\end{proof}

\subsection{The Ellipsoid Algorithm with Separation Oracles}

The ellipsoid  method of Gr{\"o}tschel,  Lov{\'a}sz and Schrijver \cite{ellipsoid,grotschel2012geometric} states that, given  a polynomial-time separation oracle,  (even if $K$ is a polytope defined by an exponential number of hyperplanes) an approximate optimal solution to the ``convex optimization problem'' can be found in polynomial time.  If $K$ is a rational polytope,  then an exact optimal solution can be found in polynomial time.
We follow the  formulation of \cite{schrijver1998theory} in stating these results.
\begin{definition} [\cite{schrijver1998theory} Section 14.2] 
\label{def:Schi_opt}
{\bf Optimization Problem} Let $K$ be a rational polyhedron\footnote{$K$ is a  {\em Rational Polyhedron} if 
$K=\{\xx\,:\, {\bf A} \xx \le {\bf b}\}$ 
where the components of matrix $\bf A$ and vector $\bf b$ are all rational numbers.  $\mathbb{Q}$ is the set of rationals.
} 
in $\RR^m.$ Given the input $\cc \in \mathbb{Q}^m,$ conclude with one of the following:
\begin{itemize}
\item[(i)] give a vector $\xx_0 \in K$ with $\cc^T\xx_0 = \max \{\cc^T \xx\,|\, \xx \in K\}.$
\item[(ii)] give a vector $\yy_0$ in $\mbox{char.cone} \ K$ with $\cc^T \yy_0 >0.$
\item[(iii)] assert that $K$ is empty.
\end{itemize}
\end{definition}
Note that in this definition,   $\mbox{char.cone} \ K$ is the {\em characteristic cone} of $K.$ The characteristic cone of  a bounded polytope is $\{\bf 0\}$  \cite{schrijver1998theory}[Section 8.2] so,  if $K$ is a bounded nonempty polytope,  the optimization problem is to find a vector $\xx_0 \in K$ with $\cc^T\xx_0 = \max \{\cc^T \xx\,|\, \xx \in K\}.$

The relevant result of  \cite{ellipsoid,grotschel2012geometric}  is 
\begin{theorem}[\cite{schrijver1998theory} Corollary 14.1a]
\label{thm:Smain}
There exists an algorithm $\mbox{\tt ELL}$ such that if  $\mbox{\tt ELL}$ is given the input 
$(m,\varphi,\mbox{\tt SEP},c)$ where:\\[0.1in]
\hspace*{1cm}
\begin{minipage}{.8\textwidth}%
 $m$ and  $\varphi$ are  natural numbers and $\mbox{\tt SEP}$ is a separation oracle for some rational polyhedron $K$ in $\RR^m$, defined by linear inequalities of size at most $\varphi$ and $c \in \mathbb{Q}^m,$
\end{minipage}%
\ \\[0.1in]
then $\mbox{\tt ELL}$ solves the optimization problem for $K$ for the input $c$ in time polynomially bounded by $m$, $\varphi$, the size of $c$ and the running time of $\mbox{\tt SEP}.$
\end{theorem}
%
In this statement, the {\em size} of a linear inequality is the number of bits needed to write the rational coefficients of the inequality, where the number of bits required to write rational $r = p/q,$ where $p,q$ are relatively prime integers, is $\lceil \log_2 p \rceil +\lceil \log_2 q \rceil.$

\subsection{Solving the Minimum-Cost Markov Chain problem.}
\label{sec:MCPSolution}
Combining all of the pieces, we can now prove our main result.
\begin{lemma}
\label{lem:newsol}
Given $\mbbS,$ let $\varphi$ be the maximum number of bits required to write any  transition probability  $q_i(S)$ or cost $\ell(S)$ of  a permissible state $S$.

Furthermore,  assume  some known  hyper-rectangle  
${\bf R}\subset \RR^{m-1}$ with the property that there exists   $(\xx^*,y^*) \in \MCP$ satisfying $\xx^* \in {\bf R}$ and 
$y^* = \max\{y' \,:\,   (\xx',y') \in \MCP\}$.

%

Then the minimum-cost Markov chain problem can be solved in time polynomially bounded by  $m$, $\varphi$  and $t'_S({\bf R}).$
\end{lemma}
\begin{proof}
Without loss of generality, we assume that $y^* \ge 0.$   To justify,  recall that the minimum cost Markov chain $\bfS^*=(S^*_0,\ldots,S^*_{m-1})$ has cost
$y^* = \sum_k \ell(S^*_k) \pi_k(\bfS^*) \ge \min_k \ell(S^*_k)$ where  $\pi_k(\bfS^*)\ge 0$ is the $k$'th component of the stationary distribution of $\bfS^*.$
Trivially,  if  $\ell(S^*_k)  \ge 0$ for all $k,$ then $y^*\ge 0.$

The original problem formulation does not require that $\ell(S_k) \ge 0$. But,  we can 
modify a given input  by adding  the same constant $2^\varphi$ to $\ell(S_k)$ for every $S_k \in \mbbS_k$,  $k\in [m].$ This makes every $\ell(S_k)$ non-negative so the minimum cost Markov chain  in this modified problem has non-negative cost.
Since this modification adds $2^\varphi$ to the cost of {\em every} Markov chain,  solving the modified problem solves the original problem.  Note that this modification can at most double $\varphi,$ so this does not break the statement of the lemma.
We may thus assume that  $y^* \ge 0$.


Thus 
$(\xx^*,y^*)  \in \MCP' =\MCP \cap {\bf R}'$
where 
${\bf R}' =\left\{
(\xx,y) \,:\, \xx \in  {\bf R},\, y \ge 0
\right\} \subset \RR^{m}.
$
Since   $y^*$ is bounded from above by the cost of any permissible Markov chain, $\MCP'$ is a bounded non-empty polytope.

Since $(\xx^*,y^*) \in \MCP',$     $\max\{y' \,:\,   (\xx',y') \in \MCP\}=  \max\{y' \,:\,   (\xx',y') \in \MCP'\}.$

Now consider the following separation oracle for  $\MCP'.$ Let $(\xx,y) \in \RR^m.$ 
\begin{itemize}
\item In $O(1)$ time,  first  check whether $y \ge 0.$ If  no, then $(\xx,y)\not \in \MCP'$ and $y=0$ is a separating hyperplane.
\item Otherwise,    in  $O(m)$ time, check whether $\xx \in {\bf R}.$ If no, $(\xx,y)\not \in \MCP'$ and a separating hyperplane is just the corresponding side of $\bf R$ that  $(\xx,y)$ is outside of. 
\item Otherwise,  calculate $\bfS(\xx)$ in $t_S({\bf R})$ time.  From Lemma \ref{lem:SSep}, this provides a separation oracle.
\end{itemize}


Consider solving the optimization problem on polytope $ K =  \MCP'$ with $\cc=(0,0,\ldots,0,1)$ to find $(\xx_0,y_0) \in \MCP'$  satisfying 
$$\cc^T (\xx_0,y_0) = \max\{\cc^T (\xx',y') \,:\,   (\xx',y') \in \MCP'\} = \max\{y' \,:\,   (\xx',y') \in \MCP'\} = y^*.
$$

Since  $\MCP'$ is a bounded non-empty polytope,  we can apply 
 Theorem \ref {thm:Smain}  to find such an $(\xx_0,y_0)$ in  time polynomially bounded in $m,$ $\varphi$, and $t_{\mbbS}({\bf R}).$

Applying Corollary \ref {cor:IterativeAlg} and its procedure $\mbox{\rm Prune}(\xx_0)$ 
then produces   a minimum  cost Markov-Chain in another $O(m t'_{\mbbS}({\bf R}))$ time.
The final result follows from the fact that
$t_{\mbbS}({\bf R}) \le t'_{\mbbS}({\bf R}).$
\end{proof}
 
Part 2, starting in Section \ref {sec:AIFVlag}, shows how to apply this Lemma to derive  a polynomial time algorithm for constructing minimum-cost AIFV-$m$ codes.  

We also note that \cite{DGZ2} recently applied Lemma \ref{sec:MCPSolution}  in a plug-and-play manner to derive the first polynomial time algorithms for constructing optimal AIVF codes.   AIVF codes are a multi-tree generalization of Tunstall coding  \cite{iwata2021aivf,iwata2022joint},  for which the previous constriction algorithms ran in exponential time.

%
%
%
%
%
%
%

\section{Proofs of  Lemmas  \ref{lem:geometric}   and  \ref{lem:Iterative_PruningX} }
\label{Sec:Proofs}

\subsection{Proof of Lemma \ref{lem:geometric}}

Before starting the proof, we note that (a) and the first equality of (b) are, after a change of variables,  implicit in the analysis provided in  \cite{iterative_4} of  the convergence of their iterative algorithm. The derivation there  is different than the one provided below, though,  and is missing intermediate steps that are  needed for proving our later lemmas.
 
\begin{proof} In what follows, $\QQ$ denotes $\QQ(\bfS)$, the transition matrix associated with $\bfS,$ and $\pii$ denotes $\pii(\bfS)$, its unique stationary distribution.   

To prove (a)  observe that the intersection condition
    \begin{equation}\label{eq: int1}
    y^{int} = f_{0}(\xx^{int}, S_0) = f_{1}(\xx^{int}, S_1) = \dots = f_{m-1}(\xx^{int}, S_{m-1})
    \end{equation}
    can be equivalently rewritten as
        \begin{equation}\label{eq: int2}
    \begin{bmatrix}
        y^{int} \\
        y^{int} \\
        \vdots \\
        y^{int} \\
    \end{bmatrix}
    =
    \begin{bmatrix}
        f_{0}(\xx^{int}, S_0) \\
        f_{1}(\xx^{int}, S_1) \\
        \vdots \\
        f_{m-1}(\xx^{int}, S_{m-1}) \\
    \end{bmatrix},
    \end{equation}
    where the right-hand side of (\ref{eq: int2}) can be expanded into
    $$
    \begin{bmatrix}
        \ell(S_0) \\
        \ell(S_1) \\
        \vdots \\
        \ell(S_{m-1}) \\
    \end{bmatrix}
    +
    \begin{bmatrix}
        q_0(S_0) & q_1(S_0) & \dots & q_{m-1}(S_0) \\
        q_0(S_1) & q_1(S_1) & \dots & q_{m-1}(S_1) \\
        \vdots & \vdots & \ddots & \vdots \\
        q_0(S_{m-1}) & q_1(S_{m-1}) & \dots & q_{m-1}(S_{m-1}) \\
    \end{bmatrix}
    \begin{bmatrix}
        0 \\
        x^{int}_1 \\
        \vdots \\
        x^{int}_{m-1} \\
    \end{bmatrix}
    -
    \begin{bmatrix}
        0 \\
        x^{int}_1 \\
        \vdots \\
        x^{int}_{m-1} \\
    \end{bmatrix}.
    $$
   Equation (\ref{eq: int2}) can therefore be rewritten as 
    \begin{equation}\label{eq:int2M}
    -
    \begin{bmatrix}
        \ell(S_0) \\
        \ell(S_1) \\
        \vdots \\
        \ell(S_{m-1}) \\
    \end{bmatrix}
    =
    \begin{bmatrix}
        1 & q_1(S_0) & \dots & q_{m-1}(S_0) \\
        1 & q_1(S_1) - 1 & \dots & q_{m-1}(S_1) \\
        \vdots & \vdots & \ddots & \vdots \\
        1 & q_1(S_{m-1}) & \dots & q_{m-1}(S_{m-1}) - 1 \\
    \end{bmatrix}
    \begin{bmatrix}
        -y^{int} \\
        x^{int}_1 \\
        \vdots \\
        x^{int}_{m-1} \\
    \end{bmatrix},
    \end{equation}
    where the matrix in (\ref{eq:int2M}), denoted as 
    $\MMM,$ 
    is   $\QQ$ after subtracting the identity matrix $\II$ and replacing the first column with $1$s. 
To  prove (a) it   therefore suffices to prove that 
$\MMM$
is invertible.

    
    The uniqueness of $\pii$ implies that the kernel of $\QQ-\II$ is $1$-dimensional. Applying the rank-nullity theorem, the column span of $\QQ-\II$ is $(m-1)$-dimensional. Since $(\QQ-\II)\one_m = \QQ\one_m - \II\one_m = \one_m - \one_m = \zer_m$, each column of $\QQ-\II$ is redundant, i.e., removing any column of $\QQ-\II$ does not change the column span.
    
  Next, observe that $\pii \one_m = 1 \neq 0,$ implying  that $\pii$ is not orthogonal to $\one_m$. In contrast, each vector $\vv$ in the column span of $\QQ-\II$ is of the form 
$\vv=(\QQ-\II)\xx$ for some  $\xx\in \RR^m$. Thus  $\pii \vv = \pii(\QQ-\II)\xx = (\pii\QQ-\pii\II)\xx = (\pii-\pii)\xx = \zer_m^T\xx = 0$, implying that $\pii$ is orthogonal  to $\vv$. Therefore, $\one_m$ is not in the column span of $\QQ-\II$.

Combining these two observations, replacing the first column of $\QQ-\II$ with $\one_m$  increases the rank of $\QQ-\II$ by exactly one. 
Hence, 
$\MMM$
has rank $m$. This shows invertibility, and  the proof of (a) follows.

To prove (b) and (c)  observe that
    $$
    \pii
    \begin{bmatrix}
        y^{int} \\
        y^{int} \\
        \vdots \\
        y^{int} \\
    \end{bmatrix}
    =
    \pii \one_m y^{int}
    =
    (1) y^{int}
    =
    y^{int},
    $$
    and that for all $\xx \in \RR^{m-1}$,
    \begin{align*}
        \pii
        \begin{bmatrix}
            f_{0}(\xx, S_0) \\
            f_{1}(\xx, S_1) \\
            \vdots \\
            f_{m-1}(\xx, S_{m-1}) \\
        \end{bmatrix}
        &=
        \pii
        \begin{bmatrix}
            \ell(S_0) \\
            \ell(S_1) \\
            \vdots \\
            \ell(S_{m-1}) \\
        \end{bmatrix}
        +
        \pii
        (\QQ-\II)
        \begin{bmatrix}
            0 \\
            x_1 \\
            \vdots \\
            x_{m-1} \\
        \end{bmatrix},
        \\
        &=
        \begin{bmatrix}
            \pi_0(\bfS) \\
            \pi_1(\bfS) \\
            \vdots \\
            \pi_{m-1}(\bfS) \\
        \end{bmatrix}^T
        \begin{bmatrix}
            \ell(S_0) \\
            \ell(S_1) \\
            \vdots \\
            \ell(S_{m-1}) \\
        \end{bmatrix}
        +
        {\bf 0}_m
        \begin{bmatrix}
            0 \\
            x_1 \\
            \vdots \\
            x_{m-1} \\
        \end{bmatrix},
        \\
        &=
        \sum_{k=0}^{m-1} \ell(S_k) \cdot \pi_k(\bfS),
        \\
        &=\cost(\bfS).
    \end{align*}
    Applying these observations by setting $\xx = \xx^{int}$ and  left-multiplying by  $\pii,$  
    $$
    y^{int}
    =
    \pii
    \begin{bmatrix}
        y^{int} \\
        y^{int} \\
        \vdots \\
        y^{int} \\
    \end{bmatrix}
    =
    \pii
    \begin{bmatrix}
        f_{0}(\xx^{int}, S_0) \\
        f_{1}(\xx^{int}, S_1) \\
        \vdots \\
        f_{m-1}(\xx^{int}, S_{m-1}) \\
    \end{bmatrix}
    =
\cost(\bfS).
    $$
  Applying these observations again, it follows that  
    $$\forall \xx \in \RR^{m-1},\quad 
    y^{int}
    =
cost(\bfS)
    =
    \pii
    \begin{bmatrix}
        f_{0}(\xx, S_0) \\
        f_{1}(\xx, S_1) \\
        \vdots \\
        f_{m-1}(\xx, S_{m-1}) \\
    \end{bmatrix}
    =
    \sum_{k=0}^{m-1} f_{k}(\xx, S_k) \cdot \pi_k(\bfS)
    $$
proving (b).
    Since the transition probabilities $\pi(S_k) \geq 0$ are non-negative,
    $$
\forall \xx \in \RR^{m-1},\quad 
    y^{int} \geq \min_{k \in [m]} f_{k}(\xx, S_k) \cdot \sum_{k=0}^{m-1} \pi(S_k) = \min_{k \in [m]} f_{k}(\xx, S_k),
    $$
    proving (c).
    
    (d)  follows by observing that
   $$ \forall\ \xx \in \RR^{m-1}, \quad y^{int} = \min_{k \in [m]} f_{k}(\xx, S_k)  \ge  \min_{k \in [m]} g_{k}(\xx) = h(\xx).$$
\end{proof}


\subsection{Proof of  Lemma \ref{lem:Iterative_PruningX} }

\begin{proof} To prove (a) 
set $\bfS_{|P}(\hat \xx) =\left(  S_{0|P}(\hat \xx), S_{1|P}(\hat \xx), \dots, S_{m-1|P}(\hat \xx)\right).$
Assume 
    that, $\forall k \in P,$  $g_{k|P}(\hat \xx) = f_{k}(\hat \xx, S_{k|P}(\hat \xx)) = \hat y.$  

By definition,  $S_{k|P}(\hat \xx) \in \mbbS_{k|P}$ cannot transition to any $S_j\in \mbbS_j$ where $j \not\in P$. This implies that
    $$
  \forall   k \not\in P,\  \quad   \pi_k(\bfS_{|P}(\hat\xx))= 0   
\quad \implies  \quad 
\sum_{k \in P} \pi_k(\bfS_{|P}(\hat\xx))  = 1.
    $$
 
Lemma \ref{lem:geometric} (b) then implies
\begin{eqnarray*}
\cost(\bfS_{|P}(\hat \xx)) &=& \sum_{k=0}^{m-1} f_{k}(\hat\xx, S_{k|P}(\hat \xx)) \cdot \pi_k(\bfS_{|P}(\hat \xx))\\
	&=&  \sum_{k\in P}  f_{k}(\hat\xx, S_{k|P}(\hat \xx)) \cdot \pi_k(\bfS_{|P}(\hat\xx))\\
&=&  \sum_{k\in P}  \hat y  \cdot \pi_k(\bfS_{|P}(\hat\xx)) =  \hat y.
\end{eqnarray*}

Thus, $\bfS_{|P}(\hat \xx)$ is a minimum-cost Markov chain.

To prove (b),
first note that,  we are given that  $\forall k,$  $q_0(S^*_k) >0.$ Thus, $\pi(S^*_0) > 0$ and $\{0\} \subseteq P^*$, as required.

  The Markov chain starts in state  $S^*_0$ where $0 \in P^*$. 
      We claim that if $k \in P^*$ and $j \not\in P^*$ then $q_j(S^*_k) =0.$  If not,  $\pi_j(\bfS^*) \ge \pi_k(\bfS^*) \cdot q_j(S^*_k) >0$, contradicting that
$\pi_j(\bfS^*) =0.$ Thus, $S^*_k \in \mbbS_{k|P^*}$ for all $k \in P^*$.

    Now, suppose that $P^* \subseteq P.$ Then, for all $k \in P^*$,
         \begin{equation}\label{eq:iter1}
    f_{k}(\hat \xx, S^*_k) \geq g_{k|P^*}(\hat \xx) \geq g_{k|P}(\hat \xx) \geq g_{k}(\hat \xx) \geq h(\hat \xx) = \hat y.
    \end{equation}
    On the other hand,  using Lemma \ref{lem:geometric}  (b) and the fact that $\bfS^*$ 
 is a minimum-cost Markov chain,
    \begin{equation}\label{eq:iter2}
    \hat y = \cost(\bfS^*) = \sum_{k \in P^*} f_{k}(\hat \xx, S^*_k) \cdot \pi_k(\bfS^*) \geq \hat y \sum_{k \in P^*} \pi_k(\bfS^*) = \hat y.
    \end{equation}
    The left  and right hand sides of  (\ref{eq:iter2}) are the same and $\forall k \in P^*,\, \pi_k(\bfS^*) > 0,$ so   $\forall k \in P^*,\,  f_{k}(\hat \xx, S^*_k) = \hat y$.

This in turn forces all the inequalities in (\ref{eq:iter1}) to be equalities, i.e., for all $k \in P^*$,  $g_{k|P}(\hat \xx) = \hat y.$

Thus, $P^* \subseteq \{ k \in P: g_{k|P}(\hat \xx) = \hat y \}$, proving (b).
\end{proof}

\section{A Polynomial Time Algorithm for binary  AIFV-$m$ Coding.}
\label{sec:AIFVlag}

This second part of the paper  introduces binary AIFV-$m$ codes and then  applies   Lemma \ref{lem:newsol} to find a minimum-cost code of this type  in 
time polynomial in $n$, the number of source 
words to be encoded, and $b,$ the number of bits required to state the probability of any source word.


The remainder of this section 
defines binary AIFV-$m$ codes 
and describes  how they are a
special case of the 
minimum-cost Markov chain problem. 

Section \ref{sec:subAIFV} then explains  how to apply  Lemma \ref{lem:newsol}.
%

This  first requires showing that  $\varphi$ is polynomial in $n$ and $b$,  which will be straightforward.
It   also requires identifying a hyperrectangle $\bf R$  that contains a highest point  $(\xx^*,y^*) \in \MCP$  
{\em and}
for which 
$ t'_{\mbbS}({\bf R})$ is polynomial in $n$. 
That is, 
$\forall \xx \in {\bf R},  \forall P \in \calP,$  $\bfS_{|P}(\xx)$ can be calculated in polynomial time.

As mentioned at the start of Section \ref{Sec:AlgImp},  as part of improving the running time of the iterative algorithm,   \cite{dp_2,dp_2_speedup,iterative_m,golin2022speeding} showed that, for 
${\bf R} = [0,1]^{m-1},$
$t_{\mbbS}({\bf R})$ is polynomial time.  As will be discussed in Section \ref{subsec:AppR},  the algorithms there can be easily modified to show that $t'_{\mbbS}({\bf R}) =O\left(t_{\mbbS}({\bf R}) \right)$.

Corollary \ref{cor:MCPtop2}  only tells us that there exists some  $\xx^*$ such that $(\xx^*,h(\xx^*))$ is a highest point in $\MCP$.  In order to use 
  Lemma \ref{lem:newsol}, we will need to show that there exists such an   $\xx^* \in   [0,1]^{m-1}$.
Proving this is the 
most  cumbersome  and longest part of the proof.  It   combines  a case-analysis of the tree structures of AIFV-$m$ trees with the Poincare-Miranda theorem to show that the functions 
$g_k(\xx),$ $k \in [m]$ must all mutually intersect at some point $\xx^* \in [0,1]^{m-1}.$  From Lemma \ref{lem:newopt},    $\left(  \xx^*, h\left(\xx^*\right)\right)\in \MCP$
 and is therefore the optimum point needed.  Section
 \ref{sec:AIFVtech} develops the  tools  required for this analysis.   

%
%
%
%
%
%
%
%

 
 \subsection{Background}
 \label{subsec:aifvback}

Consider  a stationary memoryless source with alphabet   $\Sigma=\{\sigma_1, \sigma_2, \dots, \sigma_n\}$ in which symbol
$\sigma_i$ is generated with  probability $p_i$. 
  Let  ${\mathcal M}= \alpha_1\, \alpha_2\, \alpha_3 \,\ldots$  be a message generated by the source.
 
 Binary compression codes    encode each $\sigma_i$ in ${\mathcal M}$ using a binary codeword.
 Huffman codes are known to be ``optimal'' such codes.  More specifically,  they are {\em Minimum Average-Cost    Binary Fixed-to-Variable Instantaneous codes}. ``Fixed-to-Variable'' denotes that the binary codewords corresponding to the different $\sigma_i$ can have different lengths.  ``Instantaneous'', that, in a bit-by-bit decoding process, the end of a codeword is recognized immediately after its last bit is scanned.  The {\em redundancy} of a code is the difference between its average-cost   and the Shannon entropy 
 $-\sum_i  p_i \log_2 p_i$ of the source.  Huffman codes  can have worst case redundancy of $1.$ 

 Huffman codes are often represented by a coding tree, with the codewords being  the leaves of the tree.
 A series of recent work
 \cite{iterative_3,iterative_4,aifv_mr,aifv_m,dp_2,iterative_m,iterative_2} introduced Binary Almost-Instantaneous Fixed-to-Variable-$m$ (AIFV-$m$) codes.  Section  \ref{subsec:CDED}   provides a complete definition as well as examples. These differ  from Huffman codes in that they use $m$ different coding trees.
 Furthermore,  decoding might require an {\em $m$-bit delay,} i.e., reading ahead $m$ bits before knowing that the end of a codeword has already been reached (hence ``almost''-instantaneous).  Since AIFV-$m$ codes include Huffman coding as  a special case, they are never worse than Huffman codes.   
 Their advantage is that,  at least for $m \le 5,$ they have worst-case redundancy $1/m$ \cite{aifv_m,aifv_mr},  beating Huffman coding.\footnote{Huffman coding with blocks of size $m$ will also provide worst-case redundancy of 
$1/m.$ But the block source-coding alphabet, and thus the Huffman code dictionary,  would then have size $\Theta(n^m).$ In contrast,  AIFV-$m$ codes have dictionary size $\Theta(m n).$}

Historically,  AIFV-$m$ codes were preceded by 
{\em $K$-ary Almost-Instantaneous FV  ($K$-AIFV)}  codes,  
introduced in \cite{original}.  
$K$-ary AIFV codes  used a $K$ character encoding alphabet;
for $K>2$, the procedure used $K-1$ coding trees and had  a coding delay of $1$ {bit.} For $K=2,$ it used 2 trees and had a coding delay of $2$ {bits.}

{\em Binary} AIFV-$m$ codes
were introduced later    in \cite{aifv_m}. 
These are  binary  codes that are comprised of an $m$-tuple  of binary coding trees and  have decoding delay of at most $m$ bits.   
The binary AIFV-$2$ codes of \cite{aifv_m}   are identical to the $2$-ary AIFV codes of  \cite{original}. 
Very recent work  in \cite{hashimoto2024optimal} shows that minimum cost  AIFV-$2$ codes are  minimum cost   $2$-bit delay  codes.

Constructing optimal\footnote{ {An ``optimal'' $K$-AIFV or AIFV-$m$ code   is one with minimum average encoding cost over all such codes. This will be formally specified  in   Definition \ref{def:MCTcostX}.}}
$K$-AIFV or binary  AIFV-$m$ 
codes is much more difficult than constructing  Huffman codes. 
\cite{iterative_2} described an iterative algorithm for constructing optimal binary AIFV-$2$  codes.
 \cite{iterative_m} 
 generalized  this 
and proved that,  for $m=2,3,4,5$,  under some general assumptions,  this algorithm would  terminate   and,  at termination would produce an optimal binary AIFV-$m$  code. The same  was  later proven  for $m>5$ by
\cite{iterative_3} 
and \cite{iterative_m}.  This algorithm was later generalized to solve the Minimun-Cost Markov chain problem in \cite{iterative_4} and is  the iterative algorithm referenced in  Section \ref{subsec_iter}. 

\subsection{Code Definitions,  Encoding and Decoding}
\label{subsec:CDED}
{\em Note: We assume 
 $\sum_i p_i=1$  and that each $p_i$ can be represented using $b$ bits, i.e., each   probability 
 is an integral multiple of $2^{-b}$. The running time of our algorithm will, for fixed $m,$ be polynomial in $b$ and $n$, i.e., weakly polynomial.

}

A {\em binary AIFV-$m$ code} will be an $m$-tuple  $\bfT= (T_0, T_1, \dots, T_{m-1})$ of 
$m$ binary code trees  satisfying 
Definitions \ref{def:node_typesX}  and \ref{def:codeX} below. 
Each $T_i$ contains $n$ codewords. Unlike in Huffman codes,  codewords can be  internal nodes.

\begin{figure} [t] 
    \begin{center}
         \includegraphics[width=4.5in]{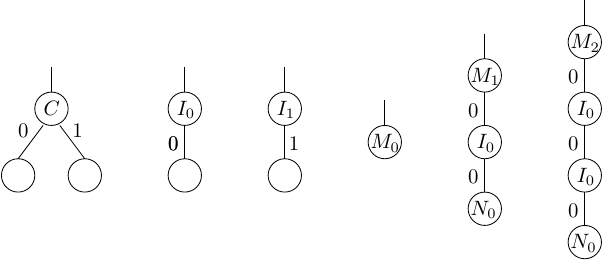}
         \end{center}
    \caption{Node types in a binary AIFV-$3$ code tree: complete node ($C$), \slave-$0$ and \slave-$1$ nodes ($I_0$, $I_1$), master nodes of degrees $0, 1, 2$ ($M_0$, $M_1$, $M_2,$ where $M_0$ is a leaf) and non-\slave-$0$ nodes (${N_0}$). The $ {N_0}$ nodes can be complete, master, or \slave-$1$ nodes, depending upon their location.} 
    \label{fig:node_typesX}
\end{figure}

\begin{definition}[Node Types in a Binary AIFV-$m$ Code  \cite{aifv_m}]
    \label{def:node_typesX}
    {\em Figure \ref{fig:node_typesX}.}
    Edges in an AIFV-$m$ code tree are labelled  as $0$-edges or $1$-edges.
 If node $v$ is connected to its child node $u$ via a $0$-edge ($1$-edge)   then $u$ is $v$'s $0$-child ($1$-child).
 We will often identify a node interchangeably with its associated (code)word.  For example, $0^2 1 0$ is the node reached by following the edges $0,0,1,0$ down from the root.
  Following 
\cite{aifv_m}, the nodes in AIFV-$m$ code trees can be classified as being exactly one of  3 different types:
    \begin{itemize}
   \item {\em Complete Nodes.} A complete node has two children: a  $0$-child and a  $1$-child.
   A complete node has no  source symbol assigned to it.
        \item {\em \Slave  \ Nodes.} {\em \small (Some earlier papers called these {\em slave} nodes.)} An  \slave \ node has no  source symbol assigned to it and has exactly one child.
        An \slave \ node with a $0$-child  is  called an \slave-$0$ node;  with  a $1$-child is  called an \slave-$1$ node.
        \item {\em Master Nodes.} A master node has an assigned source symbol and at most one child node.  Master nodes have   associated  {\em degrees}:
        \begin{itemize}  \item a master node of degree $k = 0$ is a leaf.
        \item a master node  $v$ of degree $k \geq 1$  is connected to its unique child by a  $0$-edge.  Furthermore,
 it has exactly $k$ consecutive \slave-$0$ nodes as its direct   descendants, i.e.,  $v\,0^t$ for $0 < t \le k$ are \slave-$0$ nodes while $v\,0^{k+1}$ is not an \slave-$0$ node.
        \end{itemize}
    \end{itemize}
\end{definition}

\begin{figure} [t] 
 \begin{center}
 \includegraphics[width=4.3in]{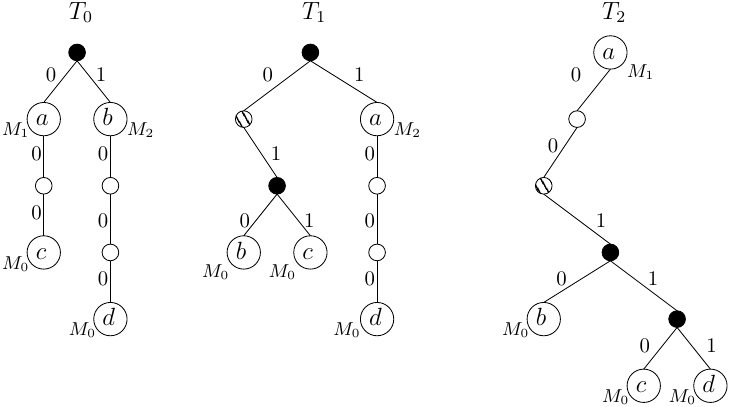}
 \end{center}
    \caption{Example binary AIFV-$3$ code for source alphabet $\left\{ a,  b,  c,  d\right\}.$ The small filled nodes are complete nodes; the small striped nodes are \slave-$1$ nodes and the small empty nodes are \slave-$0$ ones.  The  large nodes are master nodes with their assigned source symbols.  They are labelled to their sides as $M_i$ nodes, indicating that they are master-$i$ nodes.  $M_0$ nodes are leaves.
    Note that $T_2$ encodes $a$, which is at its  root, with an empty string!}
    \label{fig:example_aifv_3X}
\end{figure}

Binary AIFV-$m$ codes are now defined as follows:
\begin{definition}[Binary AIFV-$m$ Codes \cite{aifv_m}]
    \label{def:codeX}
    {\em See Figure \ref{fig:example_aifv_3X}.} Let $m \geq 2$ be a positive integer. A binary AIFV-$m$ code is an ordered $m$-tuple of $m$ code trees $(T_0, T_1, \dots, T_{m-1})$ satisfying the following conditions:
    \begin{enumerate}
        \item Every node in each code tree is either a complete node, an \slave \ node, or a master node of degree $k$ where $0 \leq k < m$.
        \item For $k \geq 1$,   code tree $T_k$ has an \slave-$1$ node connected to the root by exactly $k$  $0$-edges, i.e., the node $0^k$ is an \slave-$1$ node.
    \end{enumerate}
\end{definition}
\par\noindent{\em Consequences of the Definitions:}
\begin{enumerate}
\item[(a)] Every leaf 
is  a master node of degree $0$.  In particular,  this implies that every code tree contains at least one master node of degree $0.$ 
\item[(b)]   Definition \ref{def:codeX}, and in particular Condition (2),  result in  unique decodability (proven in  \cite{aifv_m}).
\item[(c)]  For $k \not=1,$ the root of a $T_k$ tree  is permitted to be  a master node. If a root is a master node, the associated codeword is the empty string (Figure
\ref {fig:example_aifv_3X})! The root of a $T_1$ tree  cannot be a master node.
\item[(d)]  The root of a $T_k$ tree may be an \slave-$0$ node. 
\\(This is sometimes required to maintain Definition \ref{def:codeX},   Condition (2).)
\item[(e)] For $k >0,$  every $T_k$ tree must contain  at least one \slave-$1$ node, the node $0^k.$
A  $T_0$  tree might not contain any \slave-$1$ node.\\
For $k >0,$ the root of a $T_k$ tree cannot be a   \slave-$1$ node.  The root of a $T_0$ tree is permitted to be an \slave-$1$ node (but see  Lemma \ref{lem:node_restrictionX}).
\end{enumerate}

\begin{figure}[t]  
    \centering
    \begin{tikzpicture}[thick]
        \node (T1) at (0,0) {$T_0$};
        \node (T2) at (2,0) {$T_0$};
        \node (T3) at (4,0) {$T_2$};
        \node (T4) at (6,0) {$T_1$};
        \node (s1) at (0,1) {$ c$\vphantom{$ {abcd}$}};
        \node (s2) at (2,1) {$ b$\vphantom{$ {abcd}$}};
        \node (s3) at (4,1) {$ a$\vphantom{$ {abcd}$}};
        \node (s4) at (6,1) {$ b$\vphantom{$ {abcd}$}};
        \node (c1) at (0,-1) {$000$};
        \node (c2) at (2,-1) {$1$};
        \node (c3) at (4,-1) {$\epsilon$};
        \node (c4) at (6,-1) {$010$};
        \draw[->] (T1) -- (T2);
        \draw[->] (T2) -- (T3);
        \draw[->] (T3) -- (T4);
        \draw[->] (s1) -- (T1);
        \draw[->] (s2) -- (T2);
        \draw[->] (s3) -- (T3);
        \draw[->] (s4) -- (T4);
        \draw[->] (T1) -- (c1);
        \draw[->] (T2) -- (c2);
        \draw[->] (T3) -- (c3);
        \draw[->] (T4) -- (c4);
    \end{tikzpicture}
    \caption{Encoding $ c  b  a  b$ using the AIFV-$3$ code in Figure \ref{fig:example_aifv_3X}.
      $c$ is encoded  via  $T_0$ as  ``$000$'' using a degree-$0$ master node (leaf).
    Thus the first $b$ is encoded via $T_0$ as ``$1$'' using a degree-$2$ master node.
    Then  $a$ is encoded via $T_2$ as the empty string ($\epsilon$)  using a degree-$1$ master node.
    Then the second $b$ is encoded via $T_1$ as ``$010$'' using a degree-$0$ master node. 
    Thus, $ c  b  a  b$ is encoded as $000\, 1\, \epsilon\, 010,$ i.e.,   $0001010.$
     }
 %
  %
    \label{fig:example_encodingX}
\end{figure}
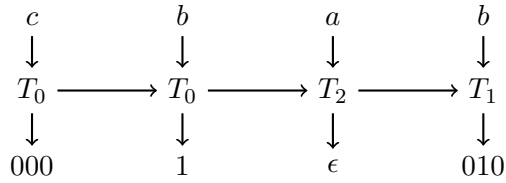

\begin{figure}  [t]  
    \centering
    \begin{tikzpicture}[thick]
        \node (T1) at (0,0) {$T_0$};
        \node (T2) at (2,0) {$T_0$};
        \node (T3) at (4,0) {$T_2$};
        \node (T4) at (6,0) {$T_1$};
        \node (c1) at (0,1) {$0001010$};
        \node (c2) at (2,1) {\sout{$000$}$1010$};
        \node (c3) at (4,1) {\sout{$0001$}$010$};
        \node (c4) at (6,1) {\sout{$0001$}$010$};
        \node (s1) at (0,-1) {$ c$\vphantom{$ {abcd}$}};
        \node (s2) at (2,-1) {$ b$\vphantom{$ {abcd}$}};
        \node (s3) at (4,-1) {$ a$\vphantom{$ {abcd}$}};
        \node (s4) at (6,-1) {$ b$\vphantom{$ {abcd}$}};
        \draw[->] (T1) -- (T2);
        \draw[->] (T2) -- (T3);
        \draw[->] (T3) -- (T4);
        \draw[->] (c1) -- (T1);
        \draw[->] (c2) -- (T2);
        \draw[->] (c3) -- (T3);
        \draw[->] (c4) -- (T4);
        \draw[->] (T1) -- (s1);
        \draw[->] (T2) -- (s2);
        \draw[->] (T3) -- (s3);
        \draw[->] (T4) -- (s4);
    \end{tikzpicture}
    \caption{Decoding $0001010$ using the binary AIFV-$3$ code in Figure \ref{fig:example_aifv_3X}.}
    \label{fig:example_decodingX}
\end{figure}
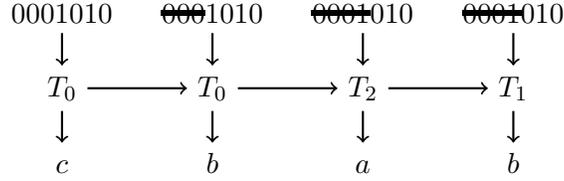

We now describe the encoding and decoding procedures. These are illustrated in Figures \ref{fig:example_encodingX} and \ref{fig:example_decodingX}.
\begin{procedure}[Encoding of a Binary AIFV-$m$ Code]
    A source sequence  ${\mathcal M}= \alpha_1 \alpha_2 \dots$ is encoded as follows: Set $T = T_0$ and $i=1.$
  \begin{tabbing}
  1.  \= Encode $\alpha_i$ using $T$\\
  2. \> Let $k$ be the index such that $\alpha_i$ is encoded using a degree-$k$ master node in $T$\\
  3. \> Set $T= T_k;$  $i=i+1$\\
  4. \> Goto line 1
  \end{tabbing}
%
\end{procedure}

\begin{procedure}[Decoding of a Binary AIFV-$m$ Code]
    Let $ \beta$ be a binary string that is the encoded message.
    Set $T = T_0$ and $i=1.$
  \begin{tabbing}
  1.  \= Let $\beta_i$ be \= longest  prefix of $\beta$ that corresponds to a path from the root of $T$ \\
 				\>\>  to some master node $M$  in $T$  \\
  2. \>  Let $k$ be the degree of $M$ (as a master node) in $T$\\
  3.\> Set $\alpha_i$ to be the source symbol assigned to $\beta_i$ in $T$\\
  4.\> Remove $\beta_i$ from the start of $\beta.$\\
  5. \> Set $T= T_k;$  $i=i+1$\\
  6. \> Goto line 1
  \end{tabbing}
%
%
%
\end{procedure}
In order to identify $\beta_i$,  line 1 might require reading a few bits after the end of $\beta_i$. The number of extra bits that must be read is known as the {\em delay}.


\begin{theorem}[\cite{aifv_m}, Theorem 3]
    Binary AIFV-$m$ codes are uniquely decodable with delay at most $m$.
\end{theorem}

\subsection{The cost of AIFV-$m$ codes}
\label{subsec:tree cost}

\begin{figure}[t]

\begingroup
    \fontsize{10pt}{10pt}\selectfont
$$\hspace*{-.7in}
\begin{array}{||cc||ccc|ccc|ccc||}\hline
i & \sigma_i & \ell(T_0,\sigma_i) & d(T_0,\sigma_i) & w(T_0,\sigma_1)& \ell(T_1,\sigma_i) & d(T_1,\sigma_i) & w(T_1,\sigma_1)& \ell(T_2,\sigma_i) & d(T_2,\sigma\_i) &w(T_2,\sigma_1) \\\hline
1 & a                        &      1    &      1       &  0 &  1        &     2     &      1 &                             0               &   1       &          \epsilon                \\\hline
2 & b                        &      1   &       2      &    1&  3        &     0     &     010  &                          4                   &      0       &         0010             \\\hline
3 & c                        &       3  &        0     &     000& 3       &     0     &     011    &                      5                     &       0       &      00110               \\\hline
4 & d                        &       4  &        0       &    1000&  4    &       0    &     1000      &                  5                       &    0            &        00111        \\\hline
\end{array}
$$

$$
\begin{array}{|c|ccc|} \hline
{j}  & \MM_j(T_0)  & \MM_j(T_1) & \MM_j(T_2) \\\hline
 0 & \{3,4\}\ \ 	& \{2,3,4\}	\ \ & \{2,3,4\}	  \\
 1 & \{1\}		& \emptyset				& \{1\}	\\
 2 &\{2\}		&\{1\}				&  \emptyset \\\hline
\end{array}
\quad 
\begin{array}{|c|ccc|} \hline
{j}  & q_j(T_0)  & q_j(T_1) & q_j(T_2) \\\hline
 0 & p_3+p_4\ \ 	& p_2+p_3+p_4 	\ \ &p_2+p_3+p_4	  \\
 1 &p_1		& 0				& p_1 \\
 2 &p_2 		& p_1			&  0 \\\hline
\end{array}
$$
\endgroup

\caption{ Example of the values introduced in  Definition  \ref{def:AIFVparemX} when calculated for the AIFV-$3$ code given in Figure \ref{fig:example_aifv_3X}.  $w(T_k,\sigma_i)$ in the first table denotes the codeword for $\sigma_i$ in code tree $T_k.$ Note that the rightmost  table on the second row is the transition matrix $\QQ(\bfT)$ for the code.
}
\label{fig:exampleparamX}
\end{figure}
\begin{definition}  Let $\TT_k(m,n)$ denote the set of all  possible type-$k$ trees that can appear in a binary AIFV-$m$ code on $n$ source symbols.  $T_k$ will be used to denote a tree $T_k \in \TT_k(m,n)$.
Set $\TT(m,n) = \bigtimes_{k=0}^{m-1} \TT_k(m,n)$.  

$\bfT= (T_0,\ldots,T_{m-1}) \in \TT(m,n)$ will be called a {\em binary AIFV-$m$ code}.
\end{definition}

\begin{definition} 
\label{def:AIFVparemX}
Figure \ref{fig:exampleparamX}. 
Let $T_k \in \TT_k(m,n)$  and $\sigma_i$ be a source symbol.
\begin{itemize}
\item  $\ell(T_k, \sigma_i)$  denotes  the {\em length} of the codeword in  $T_k$ for  $\sigma_i$.
\item $d(T_k, \sigma_i)$ denotes   the {\em degree} of the master node in $T_k$ assigned to $\sigma_i$.
\item  $\ell(T_k)$  denotes the {\em average length}  of  a codeword in  $T_k$, i.e., 
$$
\ell(T_k) = \sum_{i=1}^{n} \ell(T_k, \sigma_i) \cdot p_i.
$$
\item   $\MM_j(T_k)=\{i \in \{1,2,\ldots,n \}:\, d(T_k,\sigma_i)=j\}$ is the set of indices of source nodes that are assigned master nodes of degree  $j$ in $T_k.$ 
Set 
$${\bf q}\left(T_k) = (q_0(T_k),\ldots,q_{m-1}(T_k)\right)
\quad \mbox{ where }\quad  \forall j \in [m],\,\ 
q_j(T_k) = \sum_{i \in \MM_j(T_k)} p_i.
$$
$\sum_{j \in m} q_j(T_k) =1,$ so ${\bf q}(T_k)$ is a probability distribution.
\end{itemize}
\end{definition}

If a source symbol is encoded using a  degree-$j$ master node in $T_k$, then the next source symbol will be encoded using code tree $T_j$.  Since the source is memoryless, the {\em transition probability} of encoding using code tree $T_j$ immediately after encoding using code tree $T_k$ is $q_j(T_k)$.

        \begin{figure}[t]
        \centering
        \includegraphics[width=3.6in]{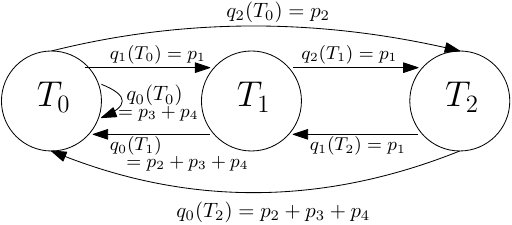}
        \caption{Markov chain corresponding to  AIFV-$3$ code in Figure \ref{fig:example_aifv_3X}. Note that $T_1$ contains no degree 1 master node, so there is no edge from $T_1$ to $T_1.$  Similarly, $T_2$ contains no degree 2 master node, so there is no edge from $T_2$ to $T_2.$}
        \label{fig:MarkovX}
        \end{figure}
This permits viewing the process  as  a  Markov chain whose states are the code trees.  Figure \ref{fig:MarkovX} illustrates an example.

From Consequence (a) following Definition \ref{def:codeX}, $\forall k\in[m],$ every $T_k \in \TT_k(m,n)$ contains at least one leaf, so $q_0(T_k) >0$. 
Thus,  as described in Section \ref{sec:MCMP},
this implies that the associated Markov chain is  a {\em unichain} whose unique recurrence class contains $T_0$   
and whose associated transition matrix $\QQ$ has a unique stationary distribution $\pii$.
Thus the Markov chain associated with any 
$\bfT \in \TT(m,n)$ is permissible.
\begin{definition}
\label{def:MCTcostX}
Let $\bfT=(T_0,\ldots,T_{m-1})\in\TT(m,n)$ be some AIFV-$m$ code,  $\QQ(\bfT)$
 be the transition matrix of the associated Markov chain and 
 $$\pii(\bfT) =(\pi_0(\bfT),\ldots,\pi_{m-1}(\bfT))$$ be $\QQ(\bfT)$'s  associated unique stationary distribution. Then the average cost of the code is the average length of an encoded symbol  in the limit, i.e., 
$$
    \cost(\bfT) = \sum_{k=0}^{m-1} \ell(T_k) \cdot \pi_k(\bfT).
    $$
\end{definition}

\begin{definition}[The Binary AIFV-$m$ Code problem]
    \label{def:global_optimization}
Construct a binary  AIFV-$m$ code 
$\bfT\in\TT(m,n)$ 
with minimum  $\cost(\bfT),$ i.e.,
$$\cost(\bfT) = \min_{\bfT'\in\TT(m,n)} \cost(\bfT').
$$
\end{definition}

This problem is {\em exactly} the minimum-cost Markov Chain problem introduced in  Section \ref{sec:MCMP} with $\mbbS_k = \TT_k(m,n)$.  
As discussed in the introduction to  Section \ref{Sec:AlgImp},   this was originally solved in exponential time by  using an iterative algorithm.


\section{Using  Lemma \ref{lem:newsol} to derive a polynomial time algorithm for binary AIFV-$m$ coding}
\label{sec:subAIFV}

Because 
the  minimum-cost binary AIFV-$m$ coding problem is a special case of the minimum-cost Markov chain problem,
 Lemma \ref{lem:newsol} can be applied  to derive a polynomial time algorithm.
In  the discussion below working through this application, $T_k$ will denote, interchangeably, both a type-$k$ tree {\em and}  a  type-$k$ state with transition probabilities 
 $q_j(T_k)$ and cost $\ell(T_k).$  For example, when writing  $f_k(\xx,T_k)$ (as in Definition \ref{def:major1}),  $T_k$ will denote the corresponding Markov chain state and not the tree.
 
Applying   Lemma \ref{lem:newsol} 
requires showing  that,  for  fixed $m,$ the $\varphi$  and $t'_{\mbbS}({\bf R})$ parameters in its  statement are polynomial in $b$ and $n$.

\subsection{Verifying that $\varphi$ is polynomial in $b,n$ for AIFV-$m$ coding}
\label{subsec:varphiAIFV}
Recall  that $\varphi$ is the maximum number of bits needed to represent the coefficients of any linear inequality defining a constraint of $\MCP.$

Showing that $\varphi$ is polynomial in $b$ and $n$ is not difficult but will  require the following bound on the height of $T_k$ trees  proven  later in
Section \ref {sec:AIFVtech}.

\medskip


\begin{lemma} \label{lem:heightX}
For all $T_k \in \TT_k(m,n),$  the height of $T_k$ is at most $n(m+1).$
\end{lemma}


We can now prove 
\begin{lemma}
\label{lem:sizeX}
${\mathbb H}$ is defined by inequalities of size  $O(m b +  \log n +  \log m)$  where $b$ is the maximum number of bits needed  to encode any of the $p_i.$
\end{lemma}
\begin{proof}
Note that the definition of ${\mathbb H}$ can be equivalently written as 
\begin{eqnarray*}
 {\mathbb H}
 &=& \left\{(\xx,y) \,|\,  (\xx\in \RR^{m-1},\, y \in \RR,\,  y \le h(\xx)\right\}\\
 &=&
 \bigcap_{k \in [m]}\ 
  \bigcap_{T_k \in \TT_{k}(m,n)} 
  \left\{
  (\xx,y) \,|\,
  \xx\in \RR^{m-1},\, y \in \RR,\,  y \le  f_{k}(\xx, T_k)
  \right\}\\
  &=&
 \bigcap_{k \in [m]}\ 
  \bigcap_{T_k \in \TT_{k}(m,n)} 
  \left\{
  (\xx,y) \,|\,
  \xx\in \RR^{m-1},\, y \in \RR,\,  y + x_k -  \sum_{j=1}^{m-1} q_j(T_k) \cdot x_j   \le  \ell(T_k)
  \right\}
\end{eqnarray*}
where we set $x_0 \equiv 0$ to provide notational consistency between the $k=0$ and $k >0$ cases.
Thus, the linear inequalities defining ${\mathbb H}$ are of the form
\begin{equation}
\label{eq:ineq}
y + x_k -  \sum_{j=1}^{m-1} q_j(T_k) \cdot x_j   \le  \ell(T_k).
\end{equation}
Since each  $p_i$ can be represented with $b$ bits,  $p_i = P_i 2^{-b}$ for
some integral $P_i \in [0,2^b].$ 
This implies that $q_j(T_k) = Q_{j,k} 2^{-b}$ for
some integral $Q_{j,k}\in [0,2^b].$ So the size of each $q_j(T_k)$ is $\le 2b.$

From Lemma  \ref{lem:heightX}
each $\ell(T_k,\sigma_i) \le n (m+1)$ for all $k,i$, so  $\ell(T_k) = \sum_{i=1}^{n} \ell(T_k, \sigma_i) \cdot p_i$ can be written as 
$L_k 2^{-b}$ for some integral $L_k \in \left[1,(n(m+1))2^b\right].$  
Thus, $\ell(T_k)$ has size
$O(b + \log n + \log m)$ 
and 
the size of every inequality (\ref{eq:ineq}) is at most $O(m b +  \log n +  \log m).$
\end{proof}
Recall that $m$ is considered fixed.  Thus $\varphi = O(b + \log n).$

\subsection{Finding Appropriate $\bf R$  and Showing that $t_{\mbbS}({\bf R})$ and $t'_{\mbbS}({\bf R})$ are  Polynomial in $b,n$ for AIFV-$m$ Coding}
\label{subsec:AppR}
Recall that $\mbbS = \TT(m,n).$  Fix  $m$ and ${\bf R} = [0,1]^{m-1}.$

As discussed at the starts  of Section \ref{Sec:AlgImp} and \ref{sec:AIFVlag},  there are dynamic programming algorithms that, for $m=2,$  give 
$t_{\mbbS}({\bf R})= O(n^3)$  \cite{dp_2_speedup} and  for $m >2,$  \
$t_{\mbbS}({\bf R})= O(n^{m+2})$ \cite{golin2022speeding}.   For $\xx \not\in {\bf R},$ the best known algorithms for calculating $\bfS(\xx)$ 
use integer linear programming and run in exponential time.

%

In deriving the  polynomial time binary search algorithm for $m=2$,  \cite{GHIEEIT2023}  proved that $\xx^* \in [0,1],$ and  could therefore  use the $O(n^3)$ DP time algorithm for $\bfS(\xx)$  as a subroutine.   We need to prove something similar for $m >2.$

The proof is quite detailed. 
The main tool used will be the following highly technical lemma whose (long)  proof  is deferred to Section \ref{sec:AIFVtech}.
   \begin{lemma}\label{lem:Pointool} Let  $m$ be fixed,  $n \ge 2^{m-1}, $  
   $\xx \in {[0,1]^{m-1}}$ and $ k  \in \{1\, \ldots,\,m-1\}.$ 
   Then
    \begin{itemize}
    \item If  $x_k=0,$   $g_{0}(\xx)  -  g_{k}(\xx) \le 0.$
    \item  If $x_k=1,$   $g_{k}(\xx)  -  g_{0}(\xx)  \le 0.$
    \end{itemize}
    \end{lemma}

The proof also needs a generalization of the intermediate-value theorem:
    \begin{theorem}[Poincar\'e-Miranda Theorem \cite{poincaremiranda}]  \label{Thm:MPXX}
        Let $f_1, f_2, \dots, f_r : [0,1]^r \to \RR$ be $r$ continuous functions of $r$ variables $x_1, x_2, \dots, x_r \in [0,1]$ such that for all indices $i$, $x_i = 0$ implies that $f_i \leq 0$ and $x_i = 1$ implies that $f_i \geq 0$. It follows that there exists a point  $(x_1, x_2, \dots, x_r)\in [0,1]^{r}$ such that 
        $\forall i,\, f_i(x_1,\ldots,x_r) = 0$.
    \end{theorem}
    
Combining  Lemma \ref{lem:Pointool}  and Theorem   \ref{Thm:MPXX} yields
 \begin{lemma}  
 \label{lem:AIFVmax}
 Let  $m$ be fixed and $n \ge 2^{m-1}.$   Then there exists  $\xx^* \in [0,1]^{m-1}$ satisfying  
 \begin{equation}
    \label{eq:gexgXX}
    g_{0}(\xx^*) = g_{1}(\xx^*) = \dots = g_{m-1}(\xx^*).
    \end{equation}
 \end{lemma}
 \begin{proof}
Set
  $r \la m-1$ and $f_k \leftarrow g_{0} - g_{k}$ for all  
  $k \in \{ 1, \dots, m-1 \}$. 
  From Lemma \ref {lem:Pointool}, if $x_k =0$ then $f_k(\xx) \le 0$ and if $x_k=1$ then $f_k(\xx) \ge 0.$   
  The Poincar\'e-Miranda   theorem then immediately  implies the existence of $\xx^* \in [0,1]^{m-1}$ such that, $\forall k\in [m], \, g_{0}(\xx^*) = g_{k}(\xx^*)$, i.e., (\ref{eq:gexgXX}).
  \end{proof}
%
%
%
%
%

Lemma \ref{lem:AIFVmax} combined with 
 Lemma \ref{lem:newopt},   immediately show that  if $n \ge 2^{m-1},$  there exists
$\xx^* \in [0,1]^{m-1}$  satisfying  $h(\xx^*) =  \max\{y' \,:\,   (\xx',y') \in \MCP\}$.

As noted earlier,   for $\xx \in {\bf R}=[0,1]^{m-1},$   there are dynamic programming algorithms for calculating $\bfS(\xx)$ in $O(n^3)$ time when $m=2$ \cite{dp_2_speedup}  and  $O(n^{m+2})$ when $m>2$ \cite{golin2022speeding}.
Thus  $t_{\mbbS}({\bf R})=O(n^3)$ for $m=2$ and  $t_{\mbbS}({\bf R})=O(n^{m+2})$ for $m>2.$

Those dynamic programming  algorithms work by building the $T_k$ trees top-down.  The status of nodes on the bottom level of the partially built  tree,  i.e.,  whether they are   complete,   \slave\,  or master nodes  of a particular degree,   is left undetermined.  One step of the dynamic programming algorithm then determines (guesses) the status of those bottom nodes and creates a new bottom level of undetermined nodes. It is easy to modify this procedure so that nodes are only assigned a status within some given
$P\in \calP.$  The modified algorithms would then calculate $\bfS_{|P}(\xx)$ in the same running time as the original algorithms, i.e.,  $O(n^3)$ time when $m=2$ and  $O(n^{m+2})$ when $m>2$.  Thus, for $\xx \in {\bf R}$,
$t'_{\mbbS}({\bf R}) = O\left( t_{\mbbS}({\bf R}) \right).$

\subsection{The Final Polynomial Time Algorithm}

Fix $m$ and set ${\bf R} = [0,1]^{m-1}$.   Since $m$ is fixed, we may assume that $n\ge 2^{m-1}.$ For smaller $n,$ the problem can be solved in $O(1)$ time by brute force.

In  the notation of   Lemma \ref{lem:newsol},  Section \ref{subsec:varphiAIFV} shows that  $\varphi= O(b +\log n)$ where $b$ is the maximum number of bits needed to encode any of the $p_i.$
Section \ref{subsec:AppR} shows that   $t'_{\mbbS}({\bf R})= O(n^3)$ when $m=2$ and $O(n^{m+2})$ when $m >2$ and there always    exists
$\xx^* \in {\bf R} $  satisfying  $h(\xx^*) =  \max\{y' \,:\,   (\xx',y') \in \MCP\}$.

Then, from  Lemma \ref{lem:newsol},   the binary AIFV-$m$ coding problem 
can be solved  in time polynomially bounded by  $\varphi$  and $t'_S({\bf R}),$ i.e., weakly polynomial in the input.

\section{Proof of  Lemma  \ref{lem:heightX}   and  Lemma \ref{lem:Pointool}}
\label{sec:AIFVtech}

The polynomial running time of the algorithm rested upon the correctness of the technical Lemmas  \ref{lem:heightX}  and \ref{lem:Pointool}.  
The proof of Lemma \ref{lem:Pointool}  requires deriving further   properties of AIFV-$m$ trees.   Lemma \ref{lem:heightX} will be a consequence of some of these derivations.

The main steps of the proof  of Lemma \ref{lem:Pointool}  are:
\begin{itemize}
\item generalize binary AIFV-$m$ code trees to  {\em extended}  binary code trees;
these are AIFV-$m$  code trees that are permitted extra leaves, unassociated with source symbols.
\item prove  Lemma \ref{lem:Pointool} for these extended trees;
\item convert this back to a  proof of the original  Lemma \ref{lem:Pointool}.
\end{itemize}
We first  introduce the concept of   extended binary code trees.
\begin{definition}[Extended Binary AIFV-$m$ Codes] Fix $k \in [m].$
        An {\em  extended binary AIFV-$m$ code  tree } $T^{ex}_{k}$ is defined exactly the same as a $T_k$ except that it is permitted to have an arbitrary number of leaves,
        i.e., master nodes of degree $0,$
         assigned the empty symbol $\epsilon.$  The  $\epsilon$-labelled leaves  are given  source probabilities $0.$\footnote{Since $T^{ex}_{k}$  satisfies Definitions \ref{def:node_typesX} and \ref{def:codeX},  just for a larger number of master nodes, it also satisfies Consequences (a)-(e) of those definitions.  This fact  will be used  in the  proof of Lemmas \ref{lem:node_restrictionX}. }
         See Figure \ref{fig:lemma1caseb}
        
        Let $\bange(T^{ex}_{k})$ denote the number of $\epsilon$-labelled leaves in $T^{ex}_{k}.$ Note that if 
        $\bange(T^{ex}_{k})=0$, then $T^{ex}_{k} \in \TT_{k}(m,n)$.
    \end{definition}

 For notational convenience, let 
 $\TT^{ex}_{k}(m,n)$, 
 $f^{ex}_{k}(\xx, T^{ex}_k)$,  
 and $g^{ex}_{k}(\xx)$ 
 respectively denote the {\em extended versions} of $\TT_{k}(m,n)$,  $f_{k}(\xx, T_k)$, and $g_{k}(\xx).$

  \begin{lemma} For  $k \in \{ 1, 2, \dots, m-1 \}$,
   (a) and (b) below always hold: 
 \label{lem:ex1}
        \begin{itemize}
            \item[(a)] There exists a function $T'_0 \,:\, \TT^{ex}_{k}(m,n) \rightarrow \TT^{ex}_{0}(m,n)$ satisfying          
            $$\ell\left(T'_0\left[T^{ex}_{k}\right] \right) = \ell\left(T^{ex}_{k}\right)
            \quad\mbox{and}\quad
            \forall j \in [m],\, q_j\left(T'_0\left[T^{ex}_{k}\right] \right)   = q_j\left(T^{ex}_{k}\right).
            $$
            \item[(b)]  There exists a function $T'_k \,:\, \TT^{ex}_{0}(m,n) \rightarrow \TT^{ex}_{k}(m,n)$ satisfying          
               $$\ell\left(T'_k\left[T^{ex}_{0}\right] \right) = \ell\left(T^{ex}_{0}\right)+1
            \quad\mbox{and}\quad
            \forall j \in [m],\, q_j\left(T'_k\left[T^{ex}_{0}\right] \right)   = q_j\left(T^{ex}_{0}\right).
            $$ 
        \end{itemize}
    \end{lemma}
       \begin{proof} (a) follows directly from the fact  that,   from Definition \ref{def:codeX}, 
        $\TT_{k}(m,n) \subseteq \TT_{0}(m,n)$ so
       $\TT^{ex}_{k}(m,n) \subseteq \TT^{ex}_{0}(m,n)$. Thus, simply setting
       $T'_0\left[T^{ex}_{t}\right] = T^{ex}_{k}$ satisfes the required conditions.
       
              \begin{figure}[t]
        \centering
        \includegraphics[scale=0.7]{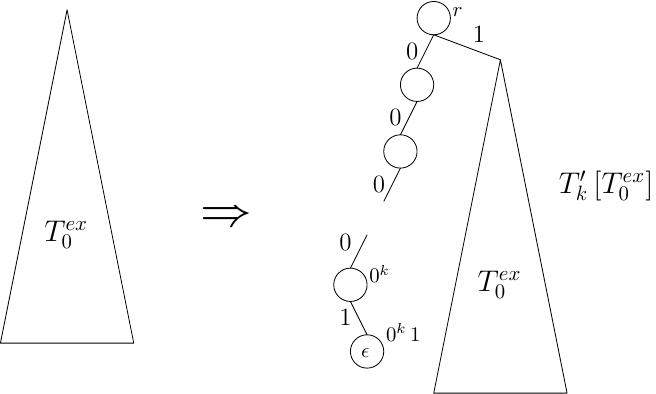}
        \caption{Illustration of Lemma \ref{lem:ex1} case (b),   describing  $T'_k\left[T^{ex}_{0}\right].$ Note that in  $T'_k\left[T^{ex}_{0}\right],$ leaf  $0^k1$ is labelled with an $\epsilon$ so the tree is an extended tree in  $\TT^{ex}_{k}(m,n)$ but not a regular tree  in  $\TT_{k}(m,n)$.}
        \label{fig:lemma1caseb}
        \end{figure}
       
       To see (b), given $T^{ex}_{0}$ consider the tree $T \in \TT^{ex}_{0}(m,n)$ whose root is complete,  with the left subtree of the root being a chain of $k-1$ \slave-$0$ nodes,  followed by one \slave-$1$ node,  and a leaf node assigned to $\epsilon$, and  with the right subtree of the root being  $T^{ex}_{0}$.  See Figure \ref{fig:lemma1caseb}.  Setting $T'_k\left[T^{ex}_{0}\right]=T$ satisfies the required conditions.
    \end{proof}
    
    This permits proving:
    \begin{lemma} Let $\xx = (x_1,\ldots,x_{m-1} )\in \RR^{m-1}.$ Then   $\forall k \in \{ 1, \dots, m-1 \},$ 
    $$\mbox{(i) }  g^{ex}_{0}(\xx)  \le g^{ex}_{k}(\xx)  + x_k
    \quad\mbox{and}\quad
    \mbox{(ii) }  g^{ex}_{k}(\xx)  + x_k  \le g^{ex}_{0}(\xx)  +1.
    $$
    \end{lemma}
    
    \begin{proof}
    Lemma \ref{lem:ex1}
    (a)  implies that for all $T^{ex}_{k} \in \TT^{ex}_{k}(m,n)$,
    \begin{eqnarray*}
        g^{ex}_{0}(\xx) 
        & \leq   &f^{ex}_{0}\left(\xx, T'_0\left[T^{ex}_{k}\right]\right)  \\
        & = &\ell\left( T'_0\left[T^{ex}_{k}\right]\right) +  \textstyle\sum_{j=1}^{m-1} q_j\left( T'_0\left[T^{ex}_{k}\right]\right)\cdot x_j \\
  	& = &\ell\left(T^{ex}_{k} \right) +  \textstyle\sum_{j=1}^{m-1} q_j\left(T^{ex}_{k} \right)\cdot x_j \\
        & = & f^{ex}_{k}(\xx, T^{ex}_{k}) + x_k.  
    \end{eqnarray*}  
    Because  this  is  true for  all $T^{ex}_{k} \in \TT^{ex}_{k}(m,n)$,  it  immediately implies (i). 
   
%

     Similarly,      Lemma \ref{lem:ex1}(b) implies that for all $T^{ex}_{0} \in \TT^{ex}_{0}(m,n)$,
      \begin{eqnarray*}
        g^{ex}_{k}(\xx) + x_k 
        &\leq & f^{ex}_{k}\left(\xx, T'_k\left[T^{ex}_{0}\right]\right)   + x_k\\
        & = &\ell\left( T'_t\left[T^{ex}_{0}\right]\right) +  \textstyle\sum_{j=1}^{m-1}  q_j\left( T'_t\left[T^{ex}_{0}\right]\right)\cdot x_j \\
   	& = & \ell\left(T^{ex}_{0} \right) +  \textstyle\sum_{j=1}^{m-1} q_j\left(T^{ex}_{0} \right)\cdot x_j  + 1\\
        & = & f^{ex}_{0}(\xx, T^{ex}_{0}) + 1.
    \end{eqnarray*}
     Because  this  is  true for  all $T^{ex}_{0} \in \TT^{ex}_{0}(m,n)$,  it  immediately implies (ii).
    \end{proof}
    
    Plugging $x_k=0$ into (i) and $x_k=1$ into (ii)  proves 
        \begin{corollary}\label{cor:PMex}  Let $\xx  \in [0,1]^{m-1}$ and $k \in \{1,\ldots,m-1\}. $Then
    \begin{itemize}
    \item If  $x_k=0,$   $g^{ex}_{0}(\xx)  -  g^{ex}_{k}(\xx) \le 0.$
    \item  If $x_k=1,$   $g^{ex}_{k}(\xx)  -  g^{ex}_{0}(\xx)  \le 0.$
    \end{itemize}
    
\end{corollary}
    Note that this is exactly Lemma \ref{lem:Pointool}  
    but written for extended binary AIFV-$m$ coding trees rather than normal ones.
    
    While  it is {\em not} necessarily true that $\forall \xx \in \RR^{m-1},$ $g^{ex}_{k}(\xx)= g_{k}(\xx)$, we can prove that if $n$ is large enough,  they coincide in the unit hypercube.
    \begin{lemma}
\label{lem:ex2} Let $n \ge  2^{m-1}.$ Then, 
        for all $k \in [m]$ and $\forall \xx \in [0,1]^{m-1}$,  $g^{ex}_{k}(\xx)= g_{k}(\xx).$ 
    \end{lemma}  
Plugging  this Lemma into  Corollary  \ref{cor:PMex}  immediately proves Lemma \ref{lem:Pointool} and thus Lemma \ref {lem:AIFVmax}.  
It therefore only remains to prove the correctness of Lemma \ref{lem:ex2}.

\subsection{Proving  Lemma \ref{lem:ex2}}

The proof of Lemma \ref{lem:ex2} is split into two parts,  The first justifies simplifying the structure of AIFV-$m$ trees. The second uses these  properties to actually prove Lemma  \ref{lem:ex2}.

\subsubsection{Further Properties of minimum-cost  $T_k$ trees}
\label{subsubsec:PropertiesX}
Definitions \ref{def:node_typesX} and \ref{def:codeX} are very loose and technically permit many scenarios,  e.g.,   the existence of more than one \slave-$1$ node in a $T_k$ tree or a chain of \slave-$0$ nodes descending from the root of a $T_0$ tree.  These scenarios will not  actually occur in trees appearing in minimum-cost codes.   The next lemma lists some of these  scenarios and justifies  ignoring them.  This will be needed in the actual proof of Lemma \ref{lem:ex2} in Section \ref{subsubsec:CompletionX}


\begin{definition}
A node $v$ is a {\em left} node of $T_k,$ if $v$ corresponds to codeword $0^t$ for some $t \ge 0$. 
\end{definition}
Note that  for $k >0,$ $T_k$ contains exactly 
$k+1$ left nodes. With the exception of  $0^k,$ which must be an \slave-$1$ node, the other left nodes can technically be  any of complete,  \slave-$0$ or master nodes.  By  definition, they cannot
be \slave-$1$ nodes.

\begin{lemma}
\label{lem:node_restrictionX}
Let $k \in [m]$ and $T'_k \in \TT^{ex}_k(m,n).$ 
Then there exists 
$T_k \in \TT^{ex}_k(m,n)$
satisfying 
\begin{equation}
\label{eq:repX}
\bange(T_k) = \bange(T'_k),\quad \ell(T_k) \le \ell(T'_k)
\quad\mbox{and}\quad
\forall j\in[m],\, q_j(T_k) = q_j(T'_k)
\end{equation}
and the 
 following five  conditions: 
\begin{itemize}
\item[(a)] The root of $T_k$ is not an \slave-$1$ node;
\item[(b)]  [Case  $k=0$] The root of $T_0$ is not an \slave-$0$ node;
\item[(c)] If $v$ is an \slave-$1$ node in $T_k,$ then the parent of $v$ 
 is not an \slave-$1$ node;
\item[(d)]  
Let $v$ be a non-root \slave-$0$ node in $T_k$.  
Additionally, if $k\neq0,$, further assume that  $v$ is not a left node.
Then the parent of $v$
 is  either a master node or an \slave-$0$ node;
\item[(e)]  If $v$ is an \slave-$1$ node in $T_k$ then $k>0$ and $v = 0^k.$\\
This implies that $v=0^k$ is the {\em unique} \slave-$1$ node in $T_k.$
\end{itemize}
\end{lemma}
    

\begin{proof} 


\medskip

 If a $T'_k$ tree does not satisfy one of conditions (a)-(d),  we first show that it can be replaced by  a  $T_k$  tree with one fewer nodes satisfying  (\ref{eq:repX}).  Since this process cannot be repeated forever, a tree satisfying all of the conditions (a)-(d) and satisfying  (\ref{eq:repX}) must exist.   
 
 Before starting, we  emphasize that none of the transformations described below adds or removes  $\epsilon$-leaves, so $\bange(T_k) = \bange(T'_k).$

\medskip


If condition (a) is not satisfied in  $T'_k,$ then,  from Consequence  (e) following Definition \ref{def:codeX}, $k=0$.  Let $r$ be the \slave-$1$ root of $T'_0$ and $v$ its child.  Create $T_0$ by removing $r$ and making $v$ the root.     Then  
 (\ref{eq:repX})  is valid (with $\ell(T_0) <  \ell(T'_0)$).


\medskip
If  condition (b) is not satisfied, let $r$ be the \slave-$0$ root of $T'_0$ and $v$ its child.  Create $T_0$ by removing $r$ and making $v$ the root.     Then again 
 (\ref{eq:repX})  is valid (with $\ell(T_0) <  \ell(T'_0)$).
Note that this argument fails for $k \not=0$ because removing the edge from $r$ to $v$ would  remove the node corresponding to word $0^k1$ and $T_k$ would then no longer be  in $\TT_k(m,n).$

\medskip

If condition (c) is not satisfied in $T'_k,$ 
let 
 $u$  be  a \slave-$1$ node in  $T'_k$ whose  child $v$ is also an \slave-$1$ node.
 Let $w$ be the unique child of $v.$   Now create $T_k$ from $T'_k$ by pointing the $1$-edge leaving $u$ to $w$ instead of $v,$  i.e., removing $v$ from the tree. Then  $\ell(T_k) \le  \ell(T'_k)$ and
(\ref{eq:repX})  is valid.
 
 \medskip

        \begin{figure}[t]
        \centering
        \includegraphics[scale=0.7]{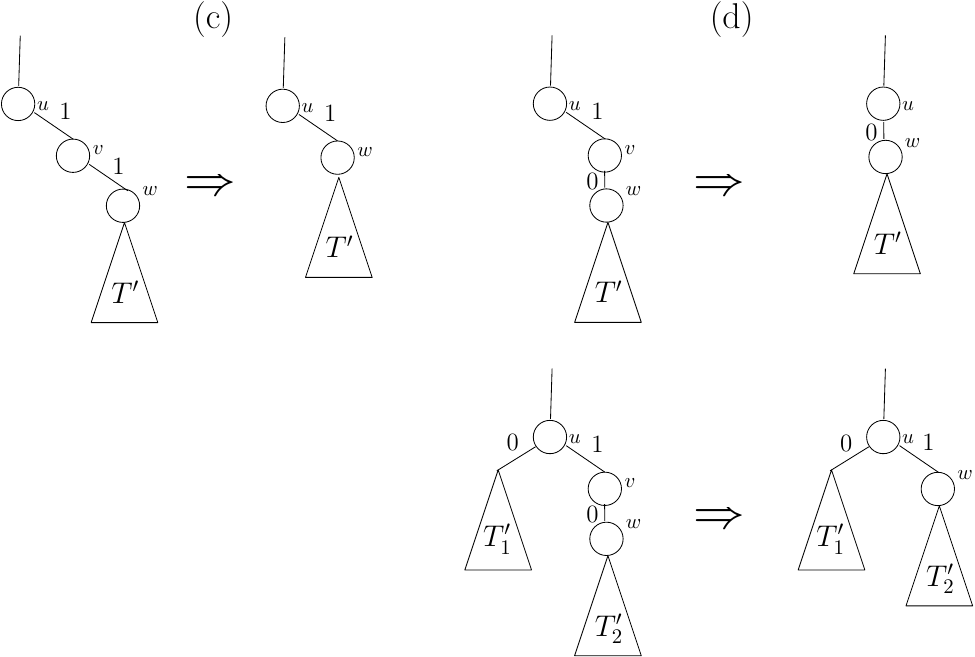}
        \caption{Illustration of transformations in cases (c) and (d) of Lemma \ref{lem:node_restrictionX}. Note that the illustration of the second subcase of (d) assumes that $v$ is the $1$-child of $u$.   The case in which $v$ is the $0$-child is symmetric.
 }
        \label{fig:NodeRestcd}
        \end{figure}

If condition (d) is not satisfied   in  $T'_k,$  let $v$ be an \slave-$0$ node in  $T'_k$ 
whose  parent $u$ is either an  \slave-$1$ node or a complete node.
  Let $w$ be the unique child of $v.$  
   Now create $T_k$ from $T'_k$ by taking  the pointer  from $u$  that was pointing to $v$ and  pointing it to $w$ instead,  i.e., again removing $v$ from $T'_k$.  
Again,  $\ell(T_k) \le  \ell(T'_k)$  and 
(\ref{eq:repX})  is valid.
Note that the condition  that 
``Additionally, if $k\neq0,$, further assume that  $v$ is not a left node'', 
ensures that Condition 2 from Definition \ref{def:codeX} is not violated.
   
   \medskip

We have shown that for any $T'_k,$ there exists a  $T_k$ satisfying (a)-(d) and  (\ref{eq:repX}).

%
%
%

\medskip
 
 Now assume that conditions (a)-(d) are satisfied in $T'_k,$ but condition 
(e) is not.  From (a), $v$ is not the root of $T_k$  so $u,$ the parent of $v,$ exists.
From (c), $u$ is not an \slave-$1$ node and from the definition of master nodes, $u$ is not a master node.  Thus $u$ is either a complete node or an \slave-$0$ node.
 Let $w$ be the unique ($1$-child) of $v.$
 From (c), $w$ cannot be an intermediate-$1$ node; from (d), $w$ cannot be an intermediate-$0$ node. So $w$ is either a complete or master node.

Now create $T_k$ from $T'_k$ by taking  the pointer  from $u$  that was pointing to $v$ and  pointing it to $w$ instead,  i.e., again removing $v$ from $T'_k.$

Note that after the transformation,  it is easy to see that 
$\ell(T_k) \le  \ell(T'_k)$ and
(\ref{eq:repX})  is valid.
Note that since $w$ is either a complete or master node,   pointing $u$ to $w$ 
does not affect the master nodes above $v.$
It only remains to show that this  pointer redirection is a permissible operation on $T_k$ trees, i.e., that $T_k$ does not violate Condition 2 from Definition \ref{def:codeX}.  


Since (e) is not satisfied, $v \neq 0^k.$ There are  two cases:

\paragraph{Case 1:  $k > 0$ and $v=0^r.$}
This is not possible for $r >k$ because $0^r \not\in T_k$.
It is also not possible for $0 < r < k,$ because in that case, $0^r$ has a $0$-child so it can not be an \slave-$1$ node.
So Case 1 cannot occur.

%
%
%


\paragraph{Case 2: $k =0$ or   $v\not=0^r$ for  any  $r>0.$} 
If $k=0$ then, trivially,   Condition 2 from Definition \ref{def:codeX} cannot be violated.   If $k \not=0$ then, since $v \neq 0^k,$  the transformation described leaves $0^k$ as an \slave-$1$ node so Condition 2 from Definition \ref{def:codeX} is still not violated.


%
%
%


%
%

      \begin{figure}[t]
        \centering
        \includegraphics[scale=0.6]{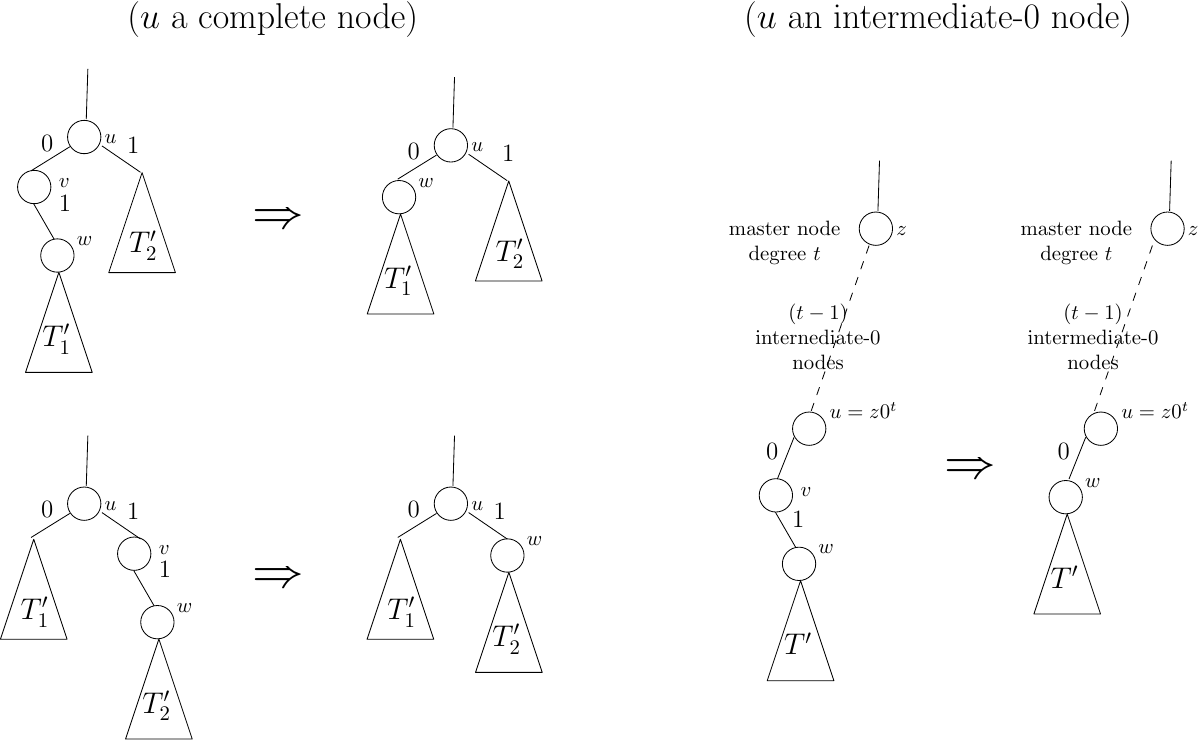}
        \caption{Illustration of transformation in case (e) of Lemma \ref{lem:node_restrictionX}.  $v$ is always an \slave-$1$ node and $w$ can be either a master node or a complete node. $z$ is a master node of degree $t$.  In the case  in which $u$ is a complete node, $v$ may be either the left or right child of $u.$  Both cases are illustrated.
 }
        \label{fig:NodeRestcd}
        \end{figure}

%
%
%
%

\medskip

%
%
%
%
This operation of removing an \slave-$1$ node can be repeated  until condition (e) is satisfied.
%
%
%
%
%
%
%
%
%
%
\end{proof}

This lemma has an immediate  corollary.
\begin{corollary}
\label{cor:limitAIFV}
There exists a minimum-cost AIFV-$m$ code $\bfT=(T_0,\ldots,T_{m-1})$ such that $\forall k \in [m],$  tree $T_k$ satisfies conditions (a)-(e) of  Lemma \ref{lem:node_restrictionX}.
\end{corollary}
\begin{proof}
Let  $\bfT'=(T'_0,\ldots,T'_{m-1})$ be a minimum cost AIFV-$m$ code. For each $k \in [m],$ let $T_k$ be the tree satisfying conditions (a)-(e) and equation (\ref{eq:repX}) and set
 $\bfT=(T_0,\ldots,T_{m-1}).$  Since $\QQ(\bfT) = \QQ(\bfT'),$  $\pii(\bfT) = \pii(\bfT').$ Since $\forall k,$  $\ell(T_k) \le  \ell(T'_k)$,
 $$
    \cost(\bfT) = \sum_{k=0}^{m-1} \ell(T_k) \cdot \pi_k(\bfT) =
       \sum_{k=0}^{m-1} \ell(T_k) \cdot \pi_k(\bfT') \le  \sum_{k=0}^{m-1} \ell(T'_k) \cdot \pi_k(\bfT') =   \cost(\bfT').
    $$
But $\bfT'$ was a minimum-cost AIFV-$m$ code so $\bfT$ must be one as well.
\end{proof}
The corollary implies that,  our algorithmic procedures, for all $k\in [m],$ may assume that  all  $T_k \in \TT_k(m,n)$ satisfy conditions  (a)-(e) of  Lemma \ref{lem:node_restrictionX}.

\medskip

This assumption permits bounding the height of all $T_k$ trees.   More specifically,   the assumption now permits proving the previously stated  Lemma \ref{lem:heightX} (that was used  in the proof of Lemma \ref {lem:sizeX}).
\smallskip

\par\noindent{\bf Lemma \ref{lem:heightX}} For all $T_k \in \TT_k(m,n),$  the height of $T_k$ is at most $n(m+1).$

\begin{proof}  Let  $T_k \in   \TT_k(m,n)$  satisfy conditions (a)-(e) of Lemma \ref{lem:node_restrictionX}.
\begin{itemize} 
\item Let $\ell\ge 1$ be the number of leaves in $T_k.$ Since all leaves are master nodes, $T_k$ contains $n-\ell$ non-leaf master nodes.
\item Every complete node in $T_k$ must contain at least one leaf  in each of its left  and right subtrees, so the number of complete nodes  in $T_k$ is at most $\ell-1.$
\item $T_k$ contains no \slave-$1$ node if $k=0$ and one \slave-$1$ node  if $k >1.$
\item If $k=0,$ each \slave-$0$ node in $T_k$ can be written as $v0^t$ for some non-leaf master node $v$ and $t \le m.$ 
If $k>0,$  all  \slave-$0$ nodes in $T_k$ with the possible exceptions of the left nodes $0^r,$  $r\le k,$ can be written as $v0^t$ for some non-leaf master node $v$ and $t \le m.$ 

So, the total number of \slave-$0$ nodes in the tree is at most $(n-\ell)m + (k+1).$
\item The total number of non-leaf nodes in the tree is then at most 
\begin{eqnarray*}
(n - \ell) + (\ell-1) + 1  + (n-\ell)m  + k+1  &=&   n + (n-\ell) m  + (k+1) \\
&\le& n + (n-1)m +  m \\
&=&  n(m+1).
\end{eqnarray*}
\end{itemize}
Thus any path from a leaf of $T_k$ to its root has length at most $n (m+1).$
\end{proof}
We note that this bound is almost tight.  Consider  a $T_0$ tree which has only one leaf and with all of the other master  nodes (including the root)  being  master nodes of degree $m.$ This tree is just a chain from the root to the unique leaf, and has   length $(n-1)(m+1).$

\subsubsection{The Actual  Proof of  Lemma \ref{lem:ex2}}
\label{subsubsec:CompletionX}
It now remains to prove Lemma \ref{lem:ex2}, i.e., 
        that if $n \ge 2^{m-1},$
 then, for all $k \in [m]$ and $\forall \xx \in [0,1]^{m-1}$,  $g^{ex}_{k}(\xx)= g_{k}(\xx).$

   \begin{proof} (of Lemma \ref{lem:ex2}.)

        Recall that $ \TT_{k}(m,n)  \subseteq  \TT^{ex}_{k}(m,n)$ so, $\forall \xx \in \RR^{m-1},$  $g^{ex}_{k}(\xx) \le g_{k}(\xx).$

Fix $ \xx \in [0,1]^{m-1}$.
        Now let  $T^{ex}_k \in \TT^{ex}_{k}(m,n)$ be a code tree satisfying
        \begin{itemize}
        \item[(i)] $f^{ex}_{k}(\xx, T^{ex}_k) = \min_{T \in \TT^{ex}_{k}(m,n)} f^{ex}_{k}(\xx, T) =  g^{ex}_{k}(\xx)$ and
        \item[(ii)]   among all trees satisfying (i), $T^{ex}_k$ is  a  tree minimizing the number of leaves assigned an  $\epsilon$.
        \end{itemize}
%
%
%
%
        Since 
        (\ref{eq:repX})  in Lemma \ref{lem:node_restrictionX} keeps $q_j(T^{ex}_k)$ the same and cannot increase $\ell(T^{ex}_k),$ we may  also assume that 
        $T^{ex}_k$ satisfies conditions (a)-(e)  of Lemma \ref {lem:node_restrictionX}.

        Since $f^{ex}_{k}(\xx, T^{ex}_k) = g^{ex}_{k}(\xx)$,
        to prove the lemma,  it  thus suffices to show that 
        $T^{ex}_k \in \TT_{k}(m,n).$
        This implies  that $g_{k}(\xx) \le g^{ex}_{k}(\xx)$ so 
$g_{k}(\xx) = g^{ex}_{k}(\xx).$ 
        
        Suppose to  the contrary that $T^{ex}_k \not\in \TT_{k}(m,n)$, i.e., $T^{ex}_k$ contains  a leaf $l_{\epsilon}$ assigned an  $\epsilon$. 
        
        Let $a_{\epsilon}$ denote the closest (lowest) non-\slave-$0$ ancestor of $l_{\epsilon}$.  
        Note that if $k=0$,  from  Lemma \ref{lem:node_restrictionX} (b), the root of $T_k$ is not a \slave-$0$ node. If $k >0,$ then, since no left node is a leaf, $\ell_\epsilon$ is not a left node,  so, one of $\ell_\epsilon$'s ancestors is a $1$-node.   Thus, in all cases, $\ell_\epsilon$ has a non  \slave-$0$ ancestor,  so $a_\epsilon$ exists.

           Now let $b_\epsilon$ be the child of $a_\epsilon$ that is the root of the subtree containing $l_\epsilon.$ 
         By the definition of $a_\epsilon,$  either $l_\epsilon = b_\epsilon$ or  
$l_\epsilon = b_\epsilon 0^{t}$ where $b_\epsilon 0^r, $ $0 \le r < t$ are   all \slave-$0$ nodes for some $t >0.$
Note again  that if $k>0,$ no left node is a leaf so, in particular neither $l_\epsilon$  or $b_\epsilon$ can be a left node.

We can therefore apply  Lemma \ref{lem:node_restrictionX} (d) to deduce that,  if $b_\epsilon$ was an \slave-$0$ node,  $a_\epsilon$ must be a master node.  Thus, if $a_\epsilon$ is a complete node or \slave-$1$ node, $b_\epsilon=l_\epsilon.$ If $a_\epsilon$ is a master node, then $b_\epsilon=a_\epsilon 0$ and $l_\epsilon = a_\epsilon 0^{t+1}$ so $a_\epsilon$ is a master node of degree $t.$


%
%
%
%
%


        Now work through the three cases: 
        \begin{itemize}
            \item[(a)] $a_{\epsilon}$ is a complete node and $b_\epsilon= l_\epsilon:$ (See Figure \ref{fig:claim1cases} (a))\\
             Let  $c_{\epsilon}$ be  the other child of   $a_{\epsilon}$ (that is not $b_\epsilon$).
Now remove   $b_\epsilon$.  If $c_\epsilon$   was  not already the $1$-child of $a_{\epsilon},$ make $c_\epsilon$  the $1$-child of $a_{\epsilon}.$

        \begin{figure}[t]
        \centering
        \includegraphics[scale=0.7]{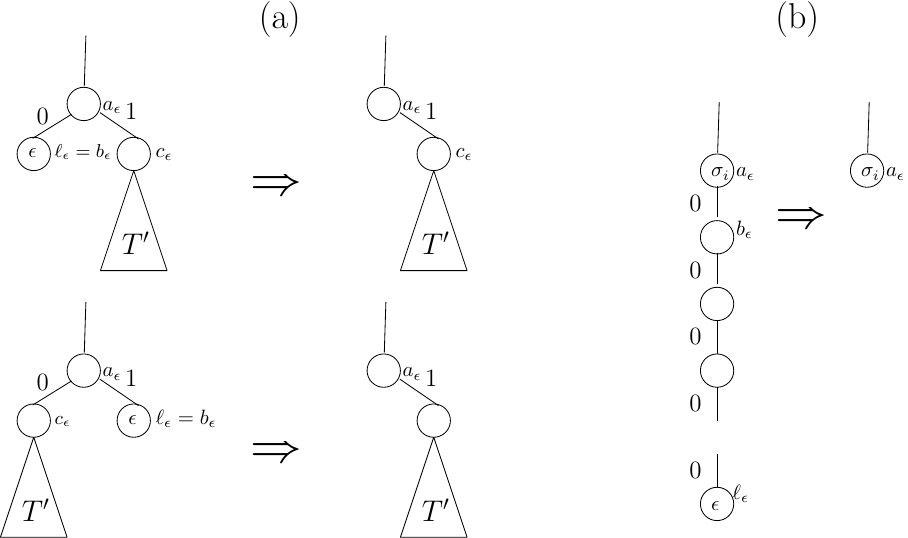}
        \caption{Illustration of  the  first two cases of  the proof of  Lemma \ref{lem:ex2}.  In case (a) $a_\epsilon$ is complete and $c_\epsilon$ can be its $1$-child or its  $0$-child.  $T'$ is the subtree rooted at $c_\epsilon.$  In case (b)  $a_\epsilon$  is a master node and $\ell_\epsilon$ is connected to it via a chain of \slave-$0$ nodes. }
        \label{fig:claim1cases}
        \end{figure}

The above transformation makes $a_e$ an \slave-$1$ node. The resulting tree remains a valid tree in $ \TT^{ex}_{k}(m,n)$ preserving the same cost $f^{ex}_{k}(\xx, T^{ex}_k)$ but reducing the number of leaves assigned to $\epsilon$ by $1$. This contradicts the minimality of $T^{ex}_k$, so this case is not possible.

            \item[(b)] $a_{\epsilon}$ is a master node of degree $t >0:$  (See Figure \ref{fig:claim1cases} (b)) \\
            Let  $\sigma_i$ be the source symbol assigned to $a_\epsilon.$  Next remove the path from  $a_{\epsilon}$ to  $l_{\epsilon},$
             converting  $a_{\epsilon}$ to a leaf, i.e., a master node of degree $0$. This reduces\footnote{This is the only location in the proof that uses  $\xx \in [0,1]^{m-1}.$} $f^{ex}_{k}(\xx, T^{ex}_k)$ by $p_i \cdot x_t \geq 0$ and the number of leaves assigned to $\epsilon$ by $1$. The resulting code tree  is still in $ \TT^{ex}_{k}(m,n)$  and 
contradicts the minimality of $T^{ex}_k$, so this case is  also not possible.


        \begin{figure}[t]
        \centering
        \includegraphics[scale=0.7]{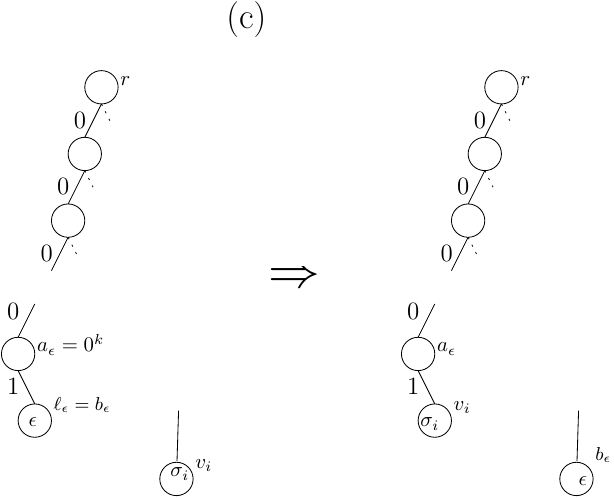}
        \caption{Illustration of case (c) of    the proof of  Lemma \ref{lem:ex2}.    $a_\epsilon= 0^k$ is an \slave-$1$ node
        and $\beta_\epsilon= 0^k1.$  From definition \ref{def:codeX} (2),  this $b_\epsilon$ must exist in every $T_k,$ so it may not be removed.  
        $v_i$ is a deepest leaf in $T_k,$ which is ``swapped'' with $b_\epsilon$.
        }
        \label{fig:claim2cases}
        \end{figure}

            \item[(c)]  $a_{\epsilon}$ is an \slave-$1$ node and
            $b_\epsilon= l_\epsilon:$     (See Figure \ref{fig:claim2cases}) \\
  %
           From Lemma \ref{lem:node_restrictionX}  (e),  $k \not=0$ and $a_{\epsilon}=0^k$,  $b_\epsilon=0^k1$.
 So,  $b_{\epsilon}$  is at depth $k+1$.

Since cases (a) and (b) cannot occur,  
$0^k1$ 
 is the {\em unique} leaf assigned an  $\epsilon.$ 
All other master nodes must be assigned some source symbol.

%
%

Let $\sigma_i$  be the deepest source symbol in the tree that is assigned to a non-left node
and  $v_i$ be the master node to which $\sigma_i$ is assigned.   Since $n\ge 2^m \ge k,$ such a $v_i$ must exist.

Since every master node in 
$T^{ex}_k$ except for $0^k1$ is assigned a  source symbol,  from 
consequence (a) following Definition \ref{def:codeX},  $v_i$ is also  a leaf.


Now swap  $b_\epsilon$ and $v_i,$  i.e., assign $\sigma_i$ to $0^k1$ and  $\epsilon$ to the node that used to be $\sigma_i.$

%

The resulting tree is  still in $ \TT^{ex}_{k}(m,n).$  
Furthermore, since the degree of all master nodes associated with source symbols  remains unchanged, $\sum_{j=1}^{m-1} q_j(T^{ex}_k) \cdot x_j$
remains unchanged.

Now consider $\ell\left(T^{ex}_k,\sigma_i\right).$

If    $\ell\left(T^{ex}_k,\sigma_i\right)>k+1$ before the swap, then 
$\ell\left(T^{ex}_k \right)$, and therefore 
$f^{ex}_{k}(\xx, T^{ex}_k),$  are  decreased by at least $p_i>0$ by the swap.
This contradicts the minimality of $T^{ex}_k$, so this is  not possible.
%
%
%

If  $\ell\left(T^{ex}_k,\sigma_i\right)=k+1$ before the swap, then $\ell\left(T^{ex}_k \right),$ and therefore 
$f^{ex}_{k}(\xx, T^{ex}_k),$  remain unchanged. Thus the new tree satisfies conditions (i) and (ii) at the beginning of the proof.  Since the original $T^{ex}_k$ satisfied   conditions (a)-(e)  of Lemma \ref {lem:node_restrictionX},  the new tree does as well.  Because of the swap, the new tree does not satisfy case (c),  so it must be in case (a) or (b).  But, as already seen,  neither of those cases can occur.  
So, 
$\ell\left(T^{ex}_k,\sigma_i\right)\not=k+1.$ 

%
%
%
%
        \end{itemize}

We have therefore proven that, if $T^{ex}_k \not\in \TT_{k}(m,n)$,  then  it must satisfy case (c)  with  
$\ell\left(T^{ex}_k,\sigma_i\right) < k+1.$   We now claim\footnote{This is the only location in the proof that uses the condition $n  \ge 2^{m-1}$.} that if $n \ge 2^k$, then 
$\ell\left(T^{ex}_k,\sigma_i\right) \ge  k+1.$  This, combined with $k \le m-1,$  would prove the lemma.

To prove the claim, let $T^{ex}_k \in \cup_{n \ge 1}\TT^{ex}_{k}(m,n)$ be some tree that maximizes the number of master nodes at  depths $\le k.$ Call such a tree a {\em $k$-maximizing tree}.

Suppose $T^{ex}_k$ contained some non-leaf master node $v$ of depth $<k$.  Transform  $v$ into a complete node by giving it a $1$-child that is a leaf.  The resulting tree has the same number of master  nodes, so it is also a $k$-maximizing tree,  but contains fewer non-leaf master nodes.  Repeating this operation yields a $k$-maximizing tree in which all of the master nodes at depth $< k$ are leaves.

Next note that  a $k$-maximizing tree  cannot contain any leaves of depth $<k$ because,  if it did, those leaves could be changed into internal nodes with two leaf  children, contradicting the definition of a $k$-maximizing tree.

We have thus seen that there is a $k$-maximizing tree $T^{ex}_k$ in which all of the master nodes at level $\le k$ are on level $k$.  But,  a binary tree has  at most $2^k$ total nodes on level $k$  and $0^k$ is not a master node in a  type-$k$ tree, so $T^{ex}_k$ contains at most $2^k-1$ master nodes on level $k.$ 
This implies that any type-$k$ tree with $\ge 2^k$ master nodes must have some (non-left) master node at depth $\ge k+1$, proving the claim and thus the lemma. 
  \end{proof}

\section{Conclusion and Directions for Further Work}
\label{sec:newconc}
The first part of this paper  introduced the {\em minimum-cost Markov chain problem}. 
We then showed how to translate it into the problem of finding the highest point in the {\em Markov Chain Polytope} $\MCP.$
In particular, Lemma \ref{lem:newsol} in  Section \ref{sec:MCPSolution} identified the  problem specific  information that is needed to use the Ellipsoid algorithm to solve the 
 problem  in polynomial time.

This was written in a very general form so that it could be applied  to solve problems other than binary  Almost Instantaneous Fixed-to-Variable-$m$ (AIFV-$m$) coding.  For example,  recent work 
\cite{DGZ2} uses this Lemma  to derive polynomial time algorithms for AIVF-coding, a generalization of Tunstall coding 
 which previously \cite{iwata2021aivf,iwata2022joint} could only be solved in exponential time using an iterative algorithm.

Another possible problem use of Lemma \ref{lem:newsol} would  be the construction of optimal codes for finite-state noiseless channels.  \cite{iwata2022joint} recently showed how to frame this problem as a minimum-cost Markov chain one and solve it using the iterative algorithm.  This problem would definitely fit into the framework of Lemma \ref{lem:newsol}.  Unfortunately, the calculation of the corresponding $\bfS(\xx)$ in \cite{iwata2022joint}, needed by Lemma  \ref{lem:newsol} as a problem-specific separation oracle, 
 is done using Integer Linear Programming and therefore requires exponential time.  The development of a polynomial time algorithm for calculating the $\bfS(\xx)$ would, by Lemma   \ref{lem:newsol}, immediately yield polynomial time algorithms for solving the full problem.

The second part of the paper restricts itself  to binary  AIFV-$m$ codes, the original motivation for the
minimum-cost Markov Chain problem.  
 These are $m$-tuples of coding trees for lossless-coding that can be modelled as a Markov chain.
We derived properties of  AIFV-$m$ coding trees that then permitted applying  Lemma \ref{lem:newsol}.  This yielded  the first (weakly) polynomial time algorithm for constructing minimum cost binary AIFV-$m$ codes.

There are still many related open problems to resolve.  The first is to note that our polynomial time algorithm is primarily of theoretical interest and, like most Ellipsoid based algorithms,  would be difficult to implement in practice.   Can this be improved to be a practical algorithm?

The second is that our algorithm  is  only {\em weakly} polynomial,  since its running time is
dependent upon the actual sizes needed to encode the transition probabilities of the Markov Chain states   in binary. For example, in AIFV-$m$ coding, this is polynomial in the number of bits needed to encode the probabilities of the words in the source alphabet.
An obvious goal  would be, at least for AIFV-$m$ coding,  to find a {\em  strongly}  polynomial time algorithm, one whose running time only depends upon  $n.$

A third  concerns the definition of permissible Markov chains. Definition \ref{sec:MCMP} requires that  $\forall k \in [m],\,  \forall S_k \in \mbbS_k,$  $q_0(S_k) >0.$ This guarantees that every permissible  Markov Chain $\bfS$ has a unique stationary distribution.  It is also needed as   a requirement for Theorem 1 in  \cite{DGZ1}, which is used to guarantee that there exists  a distinctly-typed intersection point in $\MCP$ (Corollary \ref{cor:minmax1} only guarantees that every distinctly-typed intersection point is on or above $\MCP$).  The open question is whether     $q_0(S_k) >0$ is actually needed or whether the looser requirement that every $\bfS$ has a unique stationary distribution would suffice to guarantee that $\MCP$ contains some distinctly-typed intersection point.

A final question
would  return to the iterative algorithm
 approach of
\cite{iterative_2,iterative_m,iterative_3,iterative_4}.
Perhaps  the new geometric understanding of the problem  developed here  could  improve the performance and analysis of the iterative algorithms.

As an example,  the iterative algorithm  of     \cite{iterative_3,iterative_4,iterative_m,iterative_2} 
 can now  be interpreted as moving from point to point in the set of distinctly-typed intersection points (of associated hyperplanes), never increasing the cost of 
the associated Markov chain, finally terminating at the lowest point in this set.

  This immediately leads to a better understanding of one of the issues  with the iterative algorithms for the AIFV-$m$ problem.
  
 As noted in   Section \ref {subsec_iter}, the algorithm must solve for $\bfS(\xx)$ at every step of the algorithm. As noted in Section \ref{Sec:AlgImp}, this can be done in polynomial time if $\xx \in [0,1]^{m-1}$ but requires exponential time integer linear programming if $\xx \not\in  [0,1]^{m-1}.$ A difficulty with the iterative algorithm was that it was  not able to guarantee that at every step, or even at the final solution,  $\xx \in [0,1]^{m-1}.$  With our new better understanding of the geometry of the Markov Chain polytope for the AIFV-$m$ problem, it might now be possible to prove that the condition
 $\xx \in [0,1]^{m-1}$ always  holds during the algorithm or develop a modified iterative algorithm in which the condition always  holds.
 
 \medskip

 {\sc Acknowledgement:}  The authors would like to thank Reza~Hosseini Dolatabadi  and Arian Zamani for the generation of  Figures \ref{fig:gen1} and \ref{fig:gen2}.

\bibliographystyle{plain}
\bibliography{main}

%
%


\end{document}